\documentclass{aa}
\usepackage{graphicx}
\usepackage{times}
\usepackage{mathptm}
\usepackage{epsfig}
\usepackage{txfonts}
\usepackage[figuresright]{rotating}
\begin{document}

\title{The ATESP 5 GHz radio survey}

\subtitle{I. Source counts and spectral index properties of the faint radio 
population}

\author{I. Prandoni \inst{1}
	\and P. Parma \inst{1}
        \and M.H. Wieringa \inst{2}
	\and H.R. de Ruiter \inst{3,1}
	\and L. Gregorini \inst{4,1}
	\and A. Mignano \inst{5,1}
	\and G. Vettolani \inst{6,1}
	\and R.D. Ekers \inst{2}
	}
\offprints{I. Prandoni}
\mail{prandoni@ira.inaf.it}

\institute{INAF - Istituto di Radioastronomia, Via\,Gobetti 101, I--40129, 
Bologna, Italy \\
              \email{prandoni@ira.inaf.it}
\and CSIRO Australia Telescope National Facility, P.O. Box 76, Epping, 
NSW2121, Australia
\and INAF - Osservatorio Astronomico di Bologna, Via Ranzani 1, I--40126, 
Bologna, Italy
\and Dipartimento di Fisica, Universit\`a di Bologna, Via Irnerio 46,
I--40126, Bologna, Italy
\and Dipartimento di Astronomia, Universit\`a di Bologna, Via Ranzani 1,
I--40126, Bologna, Italy
\and INAF, Viale del Parco Mellini 84, I--00136, Roma, Italy
}

\date{Received 29 September 2005 / Accepted 4 July 2006}

\titlerunning{The ATESP 5 GHz radio survey. I. }
\authorrunning{I. Prandoni et al.}

\abstract{The nature and evolutionary properties of the faint radio population,
responsible for the steepening observed in the 1.4 GHz source counts 
below 1 milliJy, are not yet entirely clear. Radio spectral 
indices may help to constrain the origin of the radio emission in such faint 
radio sources and may be fundamental in understanding eventual links to the 
optical light.}
{We study the spectral index behaviour of sources that were found in the 
1.4 GHz ATESP survey (Prandoni et al. 2000a,b), considering that the ATESP is
one of the most extensive sub-mJy surveys existing at present.}
{Using the Australia Telescope Compact Array we observed at 5~GHz part 
of the region covered by the sub-mJy ATESP survey. 
In particular we imaged a one 
square degree area for which deep optical imaging in UBVRIJK is 
available. In this paper we present the
5~GHz survey and source catalogue,
we derive the 5~GHz source counts and we discuss the $1.4 - 5$ GHz spectral 
index properties of the ATESP sources. The analysis of the 
optical 
properties of the sample will be the subject of a following paper.} 
{The 5~GHz survey has produced a catalogue of 111 radio 
sources, complete down to a ($6\sigma$) limit 
$S_{lim}(5\; \mbox{GHz}) \sim 0.4$ mJy. 
We take advantage of the better spatial resolution at 5 GHz ($\sim 2\arcsec$ 
compared to $\sim 8\arcsec$ at 1.4 GHz) to infer radio source structures 
and sizes.
The 5~GHz source counts derived by the present sample are consistent with 
those reported in the literature, but improve significantly the 
statistics in the flux range $0.4\la S_{5\;\rm{GHz}}\la 1$ mJy.
The ATESP sources show a flattening of the $1.4 - 5$ GHz spectral index 
with decreasing flux density, which is particularly significant for the 
5~GHz selected sample. Such a flattening confirms previous results coming from 
smaller samples and is consistent with a flattening of
the 5~GHz source counts occurring at fluxes $\la 0.5$ mJy.}
{}


\maketitle

\section{Introduction}\label{sec-intr}

One of the most debated issues about the sub-milliJy radio sources, 
responsible 
for the steepening of the 1.4~GHz source counts 
(Condon 1984, Windhorst et al. 1990), is the origin 
of their radio emission. Understanding whether the dominant triggering process
is star formation or nuclear activity has important implications on the study
of the star formation/black hole accretion history with radio-selected 
samples.

However, despite the extensive work done in the last decade, the nature and 
the evolutionary properties of the faint radio population are not yet entirely 
clear. Today we know that the sub-mJy population is a mixture of 
different classes of objects (low-luminosity/high-z AGNs, star-forming 
galaxies, normal elliptical and spiral galaxies), with star-forming galaxies 
dominating the microJy ($\mu$Jy) population (see e.g. Richards et al. 1999), 
and early--type galaxies and AGNs being more important at sub--mJy and 
mJy fluxes (Gruppioni et al. 1999; Georgakakis et al. 1999; 
Magliocchetti et al. 2000; Prandoni et al. 2001b). 
On the other hand, the relative fractions of the different types of objects 
are still quite uncertain, and very little is known about the role played by 
the cosmological evolution of the different classes of objects.
Conclusions about the faint radio population are, in fact, limited by the 
incompleteness of optical identification and spectroscopy, since faint radio 
sources have usually very faint optical counterparts. Clearly very deep
($R>25$) optical follow-up for reasonably large deep 
radio samples are critical if we want to probe such radio source populations.

Also important may be multi-frequency radio observations: radio spectral 
indices may help to constrain the origin of the 
radio emission in the faint radio sources and may actually be fundamental 
for understanding eventual 
links to the optical light. This is especially true if high resolution radio 
data are available and source structures can be inferred. 

Multi--frequency radio data are available only for a few very small 
($\la 60$ sources) sub--mJy 
samples. Such studies indicate that most mJy radio sources are 
of the steep--spectrum type ($\alpha < -0.5$, assuming $S\sim \nu^{\alpha}$), 
with evidence for flattening of the spectra at lower flux densities (Donnelly 
et al. 1987; Gruppioni et al. 1997; Ciliegi et al. 2003). This flattening is 
consistent with the  presence of many flat ($\alpha > -0.5$) and/or inverted 
($\alpha>0$) spectral index sources at $\mu$Jy flux densities (Fomalont et al. 
1991; Windhorst et al 1993). On the other hand, there is still disagreement 
about the interpretation of such results. 

In the $\mu$Jy population studied by Windhorst et al. (1993) 
50\% of the sources
have intrinsic angular size $\Theta \geq 2.6 \pm 1.4$ arcsec,
corresponding to $\simeq 5 - 40$ kpc at the expected median redshift of the 
sources.
Extended (kpc--scale) steep--spectrum radio sources suggest synchrotron
emission in galactic disks, while extended flat--spectrum sources may
indicate thermal bremsstrahlung from large scale star--formation both
occasionally with opaque radio cores.
On the other hand, Donnelly et al. (1987) claim that most of the
sub--mJy blue radio
galaxies have steep radio spectra and are physically quite compact
($\leq 4$ kpc). This suggests two possible alternative mechanisms for the
radio emission: 1) a nuclear starburst occurring on a few kpc scale in the
galaxy center; 2) a non-thermal nucleus on parsec scales. Only high
resolution radio observations could decide between them.

With the aim of studying the spectral index behaviour of the faint radio 
population, we imaged at 5~GHz a one square degree area of the 
ATESP 1.4~GHz survey ($S_{\rm lim}\sim 0.5$ mJy, Prandoni et al. 2000a,b), 
for which deep ($R<25.5$) optical multi-color data is available, as part 
of an ESO public survey (e.g. Mignano et al. 2006). 
Such deep optical imaging will provide optical identification and 
photometric redshifts for most of the radio sources.

We notice that the ATESP is best
suited to study the sources populating the flux interval $0.5 - 1$ mJy,
where starburst galaxies start to enter the counts, but are not yet the 
dominant population. This means that our sample can be especially 
useful to study the issue of low-luminosity nuclear activity, possibly 
related to low efficiency accretion processes and/or radio-intermediate/quiet 
QSOs. 

The present 5~GHz observations are valuable 
because i) the higher resolution images probe the
radio source structure at small scales ($\leq 2$ arcsec) and thus we can 
hopefully distinguish between disk--scale and nuclear--scale radio emission, 
and ii) the present 5~GHz survey is the largest at sub--mJy fluxes, 
by a factor $10-100$: previous samples typically cover from $<0.01$ to $\sim 
0.1$ square degrees (e.g. Bennett et al. 1983; Fomalont et al. 1984, 1991; 
Donnelly et al. 1987; Partridge et al. 1986; Ciliegi et al. 2003). 
 
In this paper we describe the ATESP 5~GHz survey, present the 5~GHz source
catalogue and counts, and discuss the 
spectral index properties of the ATESP radio sources.
The analysis of the optical properties of the sample 
will be the subject of
a following paper (Mignano et al. in preparation).

The paper is organized as follows. In Sect.~\ref{sec-20cm} we briefly
present the 1.4 GHz ATESP survey and the multi-color optical data coming
from the ESO Deep Public Survey (DPS). Sect.~\ref{sec-obs} describes 
the ATESP 5~GHz survey observations and data reduction. 
The radio mosaics we produced are discussed in Sect.~\ref{sec-mos}, while
Sect.~\ref{sec-det} describes the source extraction and parameterization 
procedure. The 5~GHz source catalogue is presented in 
Sect~\ref{sec-5ghzsources}, together with an analysis of the source size and 
structure. The 5~GHz source counts derived from the present survey and 
the $1.4-5$ GHz spectral index properties of the ATESP sources are 
presented in Sect.~\ref{sec-results}. A summary is given in 
Sect.~\ref{sec-summary}. 

\section{The 1.4 GHz ATESP Survey and Related Optical 
Information}\label{sec-20cm}

\begin{figure*}

\vspace{-9truecm}

\resizebox{\hsize}{!}{\includegraphics{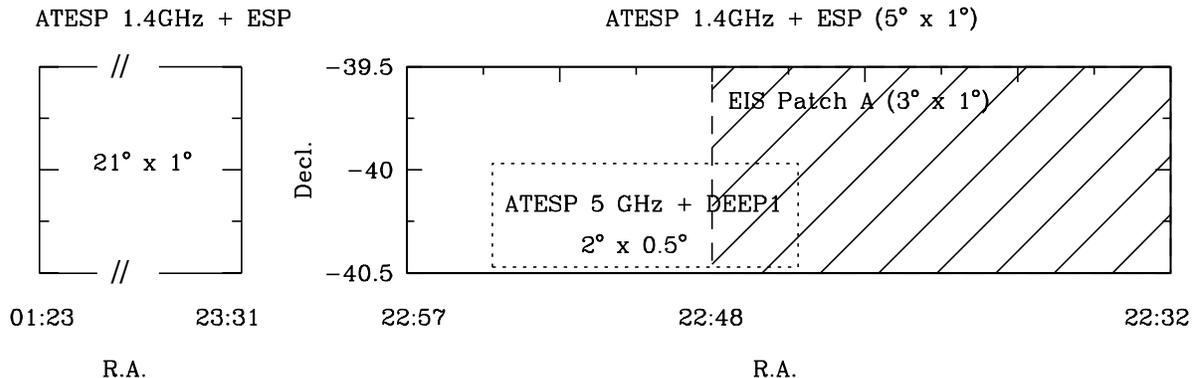}}

\vspace{-4truecm}

\caption[]{Sketch of the sky coverage of the radio (ATESP at 1.4 and 5~GHz) 
and optical (ESP, EIS, DPS) surveys relevant for this study. See text for 
details.  
\label{fig-coverage}}
\end{figure*}

The ATESP 1.4 GHz survey (Prandoni et al. \cite{Prandoni00a}) was 
carried out with the Australia Telescope Compact Array (ATCA); it 
consists of 16 mosaics with $8\arcsec \times 14\arcsec$ resolution and uniform
sensitivity ($1\sigma$ noise level $\sim 79$~$\mu$Jy), covering two 
narrow strips of $21^{\circ}\times 1^{\circ}$ and 
$5^{\circ}\times 1^{\circ}$ near the SGP, at decl. $-40^{\circ}$. 
The ATESP 1.4 GHz survey has produced 
a catalogue of 2967 radio sources, down to a flux limit ($6\sigma$) 
of $S\sim 0.5$ mJy (Prandoni et al. \cite{Prandoni00b}). 

In order to alleviate the identification work,
the area covered by the ATESP survey was chosen to overlap with
the region where Vettolani et al. (\cite{Vettolani97})
made the ESP (\emph{ESO Slice Project}) redshift survey.
They performed a
photometric and spectroscopic study of all galaxies down to $b_J 
\sim$ 19.4. The ESP survey yielded 3342 redshifts (Vettolani et al. 
\cite{Vettolani98}), to a typical depth of $z=0.1$ and a completeness level 
of 90\%. 

In the same region lies the \emph{ESO Imaging Survey} (EIS) Patch~A 
($\sim 3^{\circ}\times 1^{\circ}$ square degrees, 
centered at $22^h 40^m$, $-40^{\circ}$), mainly consisting of images in the 
I-band out of which a galaxy catalogue $95\%$ complete to $I=22.5$ has been 
extracted (Nonino et al. \cite{Nonino99}). 
This catalogue allowed us to identify $\sim 57\%$ of the 386 
ATESP sources present in that region. A first radio/optical analysis of a 
magnitude-limited sub-sample of 70 sources was presented by Prandoni et al. 
(2001b). 

More recently, a different strip of $2^{\circ} \times 0.5^{\circ}$ 
within the ATESP region at $22^h 50^m 40^s$, $-40^{\circ} 13'$
was selected for a very deep multi-color ESO public survey: the Deep Public 
Survey (DPS), which was carried out with the Wide Field Imager (WFI) at the 
2.2 m ESO telescope. 
This one square degree region is covered by 4 WFI fields (referred to 
as DEEP1a,b,c and d). In this region $UBVRI$ imaging down to very faint 
magnitudes is available (see Mignano et al. 2006): $U\sim 25$, 
$B\sim 25.8$, $V\sim 25.2$, $R\sim 25.5$, 
$I\sim 24$ ($5\sigma$, 2 arcsec aperture magnitudes). 
In addition,
DEEP1a and b have been observed in the infrared with SOFI at NTT down to 
$K_{AB}\sim 21.3$, while deeper J- and K-band images 
($J_{AB}<23.4$ and $K_{AB}<22.7$) have been taken for 
selected sub-regions (see Olsen et al., 2006) 

A sketch of the sky coverage of the surveys described in this section is given
in Fig.~\ref{fig-coverage}.

\section{The Observations}\label{sec-obs}

\subsection{Observing Strategy}\label{sec-strat}

We imaged the entire DEEP1 $2\times 0.5$ degree region at 5~GHz, since 
it has the best optical 
coverage. The area was spanned with a radio mosaic consisting of 
$21\times6$ pointings (fields) at $6'$ spacing, i.e. FWHM/$\sqrt{2}$, where 
FWHM$=10'$ is the full width at half maximum of the primary beam (see Prandoni
et al.~\cite{Prandoni00a}).
If we aim at virtually detecting all the 1.4~GHz ATESP sources with radio 
spectral index $\alpha\ge -0.7$, we need to reach a 5~GHz point source 
detection limit  of $3\sigma \simeq 0.2$ mJy.

Of course, correct spectral index determination can only be made if 
the 5 and 1.4~GHz beams 
have the same size, as severe incompleteness would result if the extra 
resolution at 5~GHz were used. 
Therefore we produced 5~GHz radio mosaics with 
the same spatial resolution as for the ATESP 1.4~GHz mosaics. 
We thus used the Compact Array in the 1.5 km 
configuration. On the other hand we also requested the 6 km antenna. 
While we want to extract the source catalogue 
from the low resolution mosaic, the longer baselines to the 6km antenna
were exploited to get additional information on the 
radio source structure (see Sect.~\ref{sec-sizes}). 

Each field (with $2\times 128$ MHz bandwidth and 10 baselines) was observed
for 98 minutes, which results in a final uniform noise level of  
$\sim 64$ $\mu$Jy. 
Therefore all $21\times 6=126$ fields could be observed in 18 
blocks of 12 hours (allowing also for calibration time). 
Care was taken to obtain 
good hour angle coverage, 
by cycling continually through the individual field during 
the observing process.
Since we wanted to observe 
the entire region also with the 6 km configuration, an
additional 12 hours were used for that purpose.

The observing log is given 
in Table~\ref{tab-log}. Note that the use of different 1.5 km arrays is not 
relevant for the present study. The two 128 MHz bands were set at 4800 and 
5056 MHz.

The flux density calibration was performed through observations of the
source PKS B1934-638, which is the standard primary calibrator for ATCA 
observations ($S=5.8$ Jy at $\nu = 4800$ MHz as revised by 
Reynolds \cite{Reynolds94}, Baars et al. \cite{Baarsetal77} flux scale). 
The phase and gain calibration was based on observations of a secondary 
calibrator (source 2254-367) selected from the ATCA calibrator list. 

\begin{table}
\caption[]{Log of the observations. \label{tab-log}}
\begin{flushleft}
\begin{tabular}{ c r c }
\hline 
\hline
\multicolumn{1}{c}{Date}
& \multicolumn{1}{c}{$t_{\mathrm{obs}}$}
& \multicolumn{1}{c}{Array}\\
\multicolumn{1}{c}{}
& \multicolumn{1}{c}{$h$}
& \multicolumn{1}{c}{}\\
\hline
 & &  \\
13/10/00 & $1\times 12$ & 6A  \\
16/11/00--19/11/00 & $3\times 12$ & 1.5B  \\
13/12/00--20/12/00 & $7\times 12$ & 1.5F  \\
02/08/01--10/08/01 & $5\times 12$ & 1.5A  \\
29/10/01--03/11/01 & $4\times 12$ & 1.5D  \\
 & &  \\
\hline
\end{tabular}
\end{flushleft}
\end{table}

\begin{table*}[t]
\caption[]{Main parameters for the 2 low resolution mosaics and average values
from the 30 full resolution mosaics. 
\label{tab-mostab}}
\begin{flushleft}
\begin{tabular}{ c r c c c c c c c r}
\hline 
\hline
\multicolumn{1}{c}{Mosaic$^a$}
& \multicolumn{1}{c}{Fields}
&\multicolumn{2}{c}{Tangent Point$^b$}
& \multicolumn{2}{c}{Synthesized Beam$^c$}
&\multicolumn{1}{c}{$S_{\mathrm{min}}$}
& \multicolumn{1}{c}{$\sigma_{\mathrm{fit}}$}
&\multicolumn{1}{c}{$<\sigma>$}\\
\multicolumn{1}{c}{fld~$x$~to~$y$}
& \multicolumn{1}{c}{$n \times m$}
&\multicolumn{1}{c}{$R.A.$}
&\multicolumn{1}{c}{$DEC.$}
& \multicolumn{1}{c}{$b_{\mathrm{min}}\times b_{\mathrm{maj}}$ ($\arcsec$)}
& \multicolumn{1}{c}{P.A. ($\degr$)}
&\multicolumn{1}{c}{mJy}
& \multicolumn{1}{c}{$\mu$Jy}
&\multicolumn{1}{c}{$\mu$Jy}\\
\hline
 & & & & & & & & & \\
 fld1to11$^d$ & $11\times 6$ & 22 47 39.57  & -40 13 00.0 & 
$7.8\times 12.8$ & $+3$  & $ -0.37 $ & 70.0 & $70.1\pm 4.2$\\
 fld10to21$^e$ & $12\times 6$ & 22 52 38.16  & -40 13 00.0 & 
$7.9\times 13.0$ & $-1$ & $ -0.33 $ & 65.2 & $64.1\pm 3.8$\\
  & & & & & & &  \\ 
 full res. mosaics$^f$ &  &  &  & 
$(2.1\times 3.5) \pm (0.1 \times 0.2)$ & $1 \pm 4 $ & 
$ -0.37 \pm 0.13$ 
& $70.1 \pm 3.6$ & $70.1\pm 4.1$\\
 & & & & & & &  \\ 
\hline
\multicolumn{9}{l}{$^a$ $x$ and $y$ refer to the first and last field columns
composing the mosaic. }\\
\multicolumn{9}{l}{$^b$ J2000 reference frame.}\\ 
\multicolumn{9}{l}{$^c$ P.A. is defined from North through East.}\\
\multicolumn{9}{l}{$^d$ Low res. mosaic overlapping the 1.4 GHz ATESP mosaic 
{\em fld05to11} (see Prandoni et al. \cite{Prandoni00a}).} \\
\multicolumn{9}{l}{$^e$ Low res. mosaic overlapping the 1.4 GHz ATESP mosaic 
{\em fld10to15} (see Prandoni et al. \cite{Prandoni00a}).}\\
\multicolumn{9}{l}{$^f$ Average values from the 30 full resolution mosaics.}
\end{tabular}
\end{flushleft}
\end{table*}

\subsection{Data Reduction}\label{sec-datared}

For the data reduction we used the 
\emph{Australia Telescope National Facility} (ATNF) release of the 
\emph{Multichannel Image Reconstruction, Image Analysis and Display} (MIRIAD) 
software package (Sault et al. \cite{Sault95}). 

Every single $12^{\mathrm{h}}$ run and each of the two 
observing bands were flagged and calibrated  
following standard procedures for ATCA observations, as described 
in Prandoni et al. (\cite{Prandoni00a}). 

Sensitivity and $u-v$ coverage were improved for each field
by merging, before imaging and cleaning, the visibilities coming from all 
the observing runs and from the two observing bands.
Imaging and deconvolution was done simultaneously for several pointings.
This is not only simpler and faster, but also produces better results, as
overlapping pointings can make use of a higher number of visibilities and 
side lobes of sources in contiguous fields can be easlily cleaned.

We produced mosaics at both {\it low} and {\it full} resolution
($\sim 10\arcsec$ and $\sim 2\arcsec$ respectively). Two low resolution
mosaics covered the entire $2\degr \times 0.5\degr$ region, while in full 
resolution
30 overlapping mosaics of 9 pointings each were produced. Final images
were obtained by cleaning after (phase only) self calibration. 

Snapshot surveys like the present one are typically affected by the 
{\it clean bias} effect: the deconvolution 
process can produce a systematic underestimation of the source fluxes, as   
consequence of the loose constraints to the cleaning algorithm due to sparse 
$u-v$ coverage (see White et al.~\cite{White97}; 
Condon et al.~\cite{Condonetal98}).
The clean bias effect has been discussed in great detail 
(Prandoni et al.~\cite{Prandoni00a}), and we repeat here only that such a 
systematic 
effect can be kept under control if cleaning is stopped well before  
the maximum residual flux has reached the theoretical noise level. 
Specifically, we set the cleaning limit at 4 times the theoretical noise, 
since 
simulations made by us show that this cut-off ensures that the clean bias 
does not affect
source fluxes.

Another systematic effect that has to be taken into account is bandwidth 
smearing. 
It is well known that at large distance from the pointing centre bandwidth 
smearing
tends to reduce the peak flux and increase the apparent source size in the 
radial direction,
such that total flux remains conserved. Also this effect has been discussed by 
us 
extensively in an earlier paper on the ATESP 1.4 GHz survey (Prandoni et al. 
2000b),
in particular in the context of radio mosaics. Considering that the passband 
width is
4 MHz, for the multichannel $32\times 4$ MHz continuum mode observations, and 
the observing 
frequency about 5000 MHz, it is easily seen from equation~(8) in Prandoni et 
al. (2000b)
that the ratio between smeared and unsmeared peak flux is between 0.9999 and 
1. 
Consequently bandwidth smearing is of no concern for our 5~GHz survey.

\section{The Radio Mosaics}\label{sec-mos}

\subsection{Production of the Mosaics}\label{sec-mosprod}

\begin{figure*}
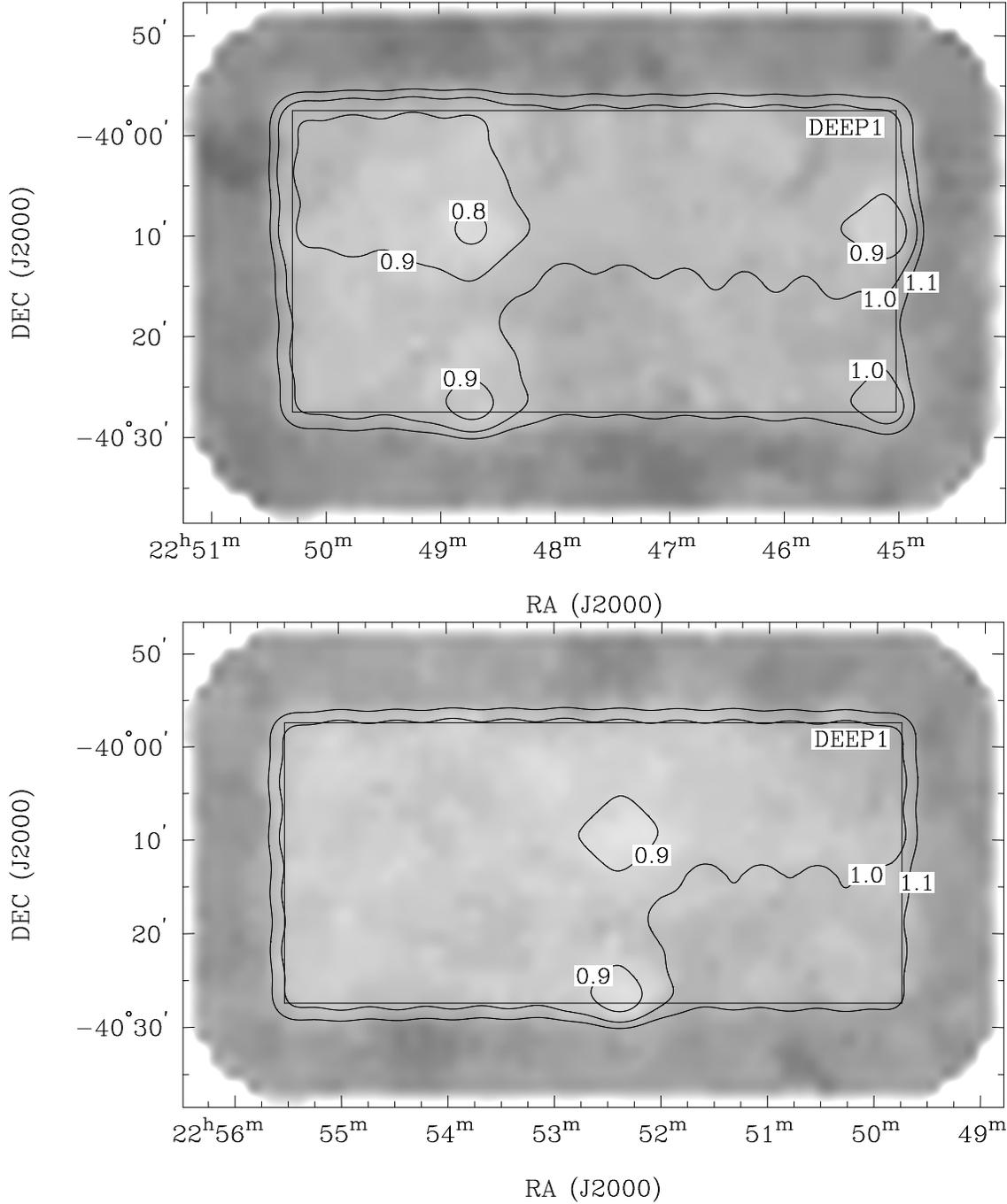

\resizebox{15cm}{!}{\includegraphics{lrmos1_noiseb.ps}}
\resizebox{15cm}{!}{\includegraphics{lrmos2_noiseb.ps}}
\caption[]{{\it Top}: 
Grey scale of the noise maps obtained from the two low resolution
mosaics: fld1to11 ({\it Top}) and fld10to21 ({\it Bottom}). 
Darker regions correspond to higher noise. The rectangular boxes indicate the 
region corresponding to the DEEP1 area, where the survey was designed to 
provide a uniform noise level. The contour images represent the theoretical
sensitivity computed by taking into account the observing time actually spent
on the mosaiced single fields: contours refer to 1.1, 
1.0, 0.9, 0.8 $\times$ the mean noise value estimated in the noise maps (see
Fig.~\ref{fig-noise}). As expected the actual noise variations trace 
reasonably well the theoretical ones.
\label{fig-noisemap}}
\end{figure*}

As mentioned before, we needed to produce mosaics at exactly the same 
resolution as the 1.4~GHz images (see Prandoni et al. \cite{Prandoni00a}),
in order to be able to compare the data at two frequencies and determine 
reliable spectral indices. 
We therefore used only the 10 baselines shorter than 3 km.
Radio maps with $520\times 520$ pixels of 2.5 arcsec were made, and combined
into two mosaics. In this way all the flux in the primary beam 
(which is 20.6 arcmin 
at 4800 MHz) is recovered. Details on the two low resolution mosaics (which
are of the order of $2000 \times 1200$ pixels. and have a small overlap) are
given in Table~\ref{tab-mostab}: we list the number of fields composing the
mosaics (columns $\times$ rows), the tangent point (sky position used for 
geometry calculations) and the restoring synthesized beam (size in arcsec and 
position angle).

Since the aim of the low resolution imaging is basically sensitivity, 
natural weighting was used in the deconvolution process. However, this 
choice may introduce some spatially correlated features and this may affect
the zero level of faint radio sources. This problem can be avoided by removing 
all baselines shorter than 60~m from the data prior to imaging and 
deconvolving.

Although this means that $\sim 3-7\%$ of the visibilities in 1.5B and 1.5F 
configurations had to be rejected, this had hardly any adverse effect on the
quality of the mosaics.
The lack of the shortest spacings would in principle lead to an increased 
insensitivity to sources larger than $90\arcsec$, but in reality less than one
source with angular size $>90\arcsec$ is expected in the area and flux range
covered by the present survey (as discussed in Sect.~\ref{sec-sizes}). 
Therefore the effect on completeness and flux densities should be minimal.

In order to assess the radio source structures full resolution images were 
produced. The $2\times 0.5$ sq.~degr. region was covered by a grid of 30
overlapping mosaics, each composed by $3\times 3$ or $3\times 2$ fields.
A size of $2060\times 2060$ pixels (with a pixel size of $0.6\arcsec$) 
for each field in the mosaics ensured complete recovery of the whole flux 
in the field of the primary beam. Some average parameters of the full 
resolution mosaics are given in Table~\ref{tab-mostab}.

\subsection{Noise Analysis of the Mosaics}\label{sec-mosanalysis}

The 5 GHz survey was designed to give uniform noise in the central 
$1.0\degr \times 0.5\degr$ regions
of the two low resolution mosaics, which together cover the area of the 
DEEP1 optical survey (see Mignano et al. 2006).
In the following our noise analysis always refers to this region. 
In column~7 of Table~\ref{tab-mostab} we list the minimum (negative) flux 
$S_{\mathrm{min}}$ found in the image, in column~8 the noise
$\sigma_{\mathrm{fit}}$ estimated as the FWHM of the flux distribution in the 
pixels in the range $\pm S_{\mathrm{min}}$, and in column~9 the noise 
$< \sigma >$ estimated as the standard deviation of the average flux in 
several source-free sub-regions of the mosaics.
$\sigma_{\mathrm{fit}}$ was used to check the presence of correlated noise,
while $< \sigma >$ gives an idea of the uniformity of the noise over the area
of the mosaic.
As expected, noise variations in general do not exceed $\sim 10\%$, with
the sole exception of one of the full resolution mosaics, due to the presence
of an $S= 27$ mJy source that could not effectively be self-calibrated.
On average we find a noise level around $\sim 70 $ $\mu$Jy; therefore a 
$3\sigma$ detection limit for a 5~GHz source at the position of a 1.4~GHz 
source is $\sim 0.21$ mJy. In all mosaics the noise is essentially Gaussian.

A more detailed analysis of the noise has been performed for the two 
low resolution mosaics, since they are used for source extraction.
The root-mean-square variations of the background were determined with the 
software package Sextractor\footnote{We used Sextractor Version 
2.2.2} (Bertin \& Arnouts 1996), which was developed for the analysis of 
optical images, but is known to work well also on radio images 
(see Bondi et al. 2003, Huynh et al. 2005). 

In computing the local background variations we used a mesh size of
$32\times32$ pixels (or equivalently $8\times 8$ beams) as this was 
empirically found to be a good compromise:  
too small mesh sizes would suffer from individual source statistics, 
while large meshes would miss large. real systematic 
variations of the background. We remark that a very similar mesh size 
proved to work well for the ATCA 1.4~GHz survey 
of the HDF South done by Huynh et al. (2005). 

The noise maps for the two low resolution mosaics
are shown in Fig.~\ref{fig-noisemap} (grey scale). 
Also shown are the expected theoretical sensitivities due
field to field variations in the actual observing time (contours). 
We notice that the measured noise is 
quite uniform in the inner part of the mosaics, and traces the expected 
sensitivity quite well.

\begin{figure}
\resizebox{\hsize}{!}{\includegraphics{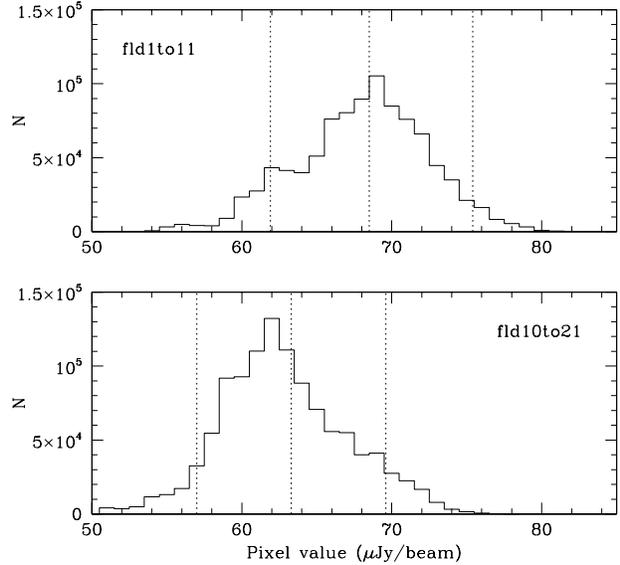}}
\caption[]{Pixel flux value distribution in the (inner part of) 
noise maps. {\it Top}: fld1to11. {\it Bottom}: fld10to21. The vertical dotted 
lines indicate the mean value and the corresponding $\pm 10\%$ variations for
each map. 
\label{fig-noise}}
\end{figure}

In Fig.~\ref{fig-noise} we plot the pixel flux value distributions in 
the noise maps. The vertical dotted lines 
indicate the mean value of the distributions ($68.5\pm 4.4$ $\mu$Jy for 
fld1to11 and $63.3\pm 4.2$ $\mu$Jy for fld10to21) and the corresponding 
$\pm 10\%$ variations. We notice that such noise values are in very good 
agreement with the average noise values obtained directly from the radio 
mosaics (see Tab.~\ref{tab-mostab}). In addition
the two distributions in Fig.~\ref{fig-noise} clearly show that noise 
variations larger than $10\%$ are very rare. 
In particular the largest noise values are
found in presence of strong radio sources, while the lowest noise values 
are due to longer observing times on some fields. 

The noise maps derived for the two low resolution mosaics have been used to
define a local signal to noise ratio, i.e. a local threshold, for the source 
detection (see Sect.~\ref{sec-det}).

\section{Source Extraction}\label{sec-det}

As for our 1.4~GHz survey we used the algorithm  
IMSAD (\emph{Image Search and Destroy}), available as part of the 
MIRIAD package for the source extraction and parameterization. The source
catalogue thus obtained is based on the low resolution mosaics exclusively.

As a first step, we extracted all sources
with peak flux $S_{\rm peak}\geq 4.5\sigma$, where $\sigma=\sigma_{fit}$ is 
the average mosaic rms flux density (see Table~\ref{tab-mostab}). 
This yielded a preliminary list of 141  
source components with $S_{\rm peak} \ga 0.3$ mJy 
in a total area of $2\times 0.5$ sq.~degr. 

All the sources were then visually inspected in order to check for
obvious failures and/or possibly poor parameterization. 
Whenever necessary the sources were re-fitted. The checking/re-fitting 
procedures adopted in this work are the same as used 
for the compilation of the 1.4 GHz ATESP source catalogue and we refer to 
Sect.~2 of Prandoni et al. (\cite{Prandoni00b}) for a full description. 
In a few cases Gaussian fits were able to provide good values for positions 
and peak flux densities, but did fail in 
determining the integrated flux densities. This happens typically at faint 
fluxes ($<10\sigma$). 
Gaussian sources with a poor determination of $S_{\rm total}$ are 
flagged in the catalogue (see Sect.~\ref{sec-cat}).    

\begin{figure}[t]
\resizebox{\hsize}{!}{\includegraphics{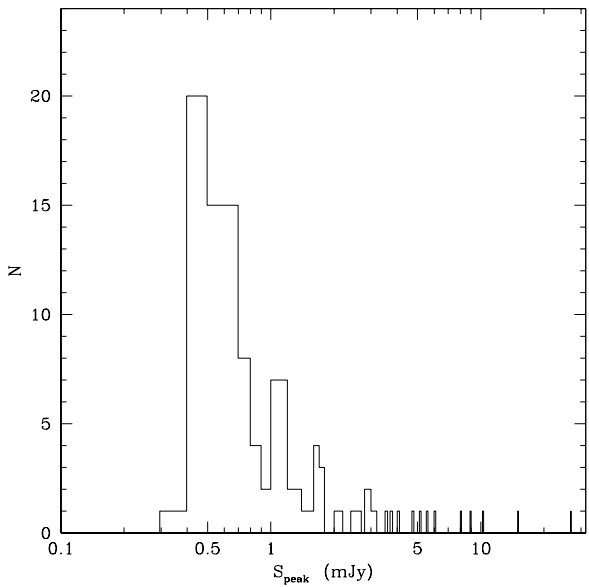}}
\caption{Peak flux density distribution of the 115 5~GHz radio sources 
(or source components) with $S_{\rm peak} \geq 6\sigma_{\rm{local}}$.
\label{fig-fluxhist}}
\end{figure}

After obtaining reliable parameters for all the sources, we used the noise 
maps described in Sect.~\ref{sec-mosanalysis} to compute the local 
signal-to-noise ratio for each source. Any source with $S_{peak} \geq 
6\sigma_{\rm{local}}$ was then included in the final list: 115 sources (or 
source components) satisfy this criterion. 
The peak flux density distribution for the final source sample 
is shown in Fig.~\ref{fig-fluxhist}. 

\begin{figure}[t]
\resizebox{\hsize}{!}{\includegraphics{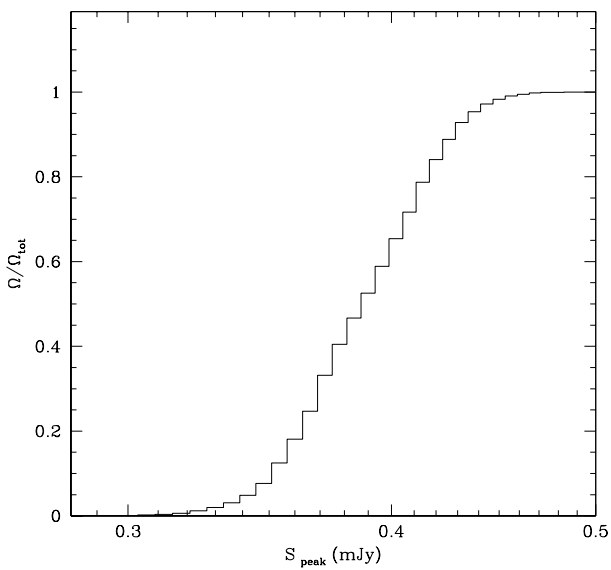}}
\caption{Visibility area of the ATESP 5 GHz survey. 
Fraction of the total area over which a source with given measured
peak flux density can be detected. 
\label{fig-areaflux}}
\end{figure}

Fig.~\ref{fig-areaflux} shows the so-called {\it visibility area}
of the ATESP 5 GHz survey, 
i.e. the fraction of the total area covered by the survey, over which a source 
with given $S_{peak}$ satisfies the $S_{\rm{peak}} \geq 6\sigma_{\rm{local}}$ 
criterion. We notice that the visibility area increases quite rapidly with flux
and becomes equal to 1 at $S_{\rm peak} \sim 0.5$ mJy. 

\subsection{Multiple and Non-Gaussian Sources}\label{sec-mult}

When we search for possible multiple-component sources, we 
find that the nearest neighbour pair separations are always $d\ga 40\arcsec$,
with the exception of four cases, where pair separations are $13.1\arcsec$, 
$18.3\arcsec$, $18.4\arcsec$ and $20.0\arcsec$ respectively. 
We decided to catalogue these 4 pairs with $d\la 20\arcsec$ as double sources. 
We notice that three of them show clear signs of 
association and are the result of the splitting performed by the source 
fitting procedure itself. Moreover, they are listed as double sources also 
in the ATESP 1.4 GHz catalogue (Prandoni et al. 2000b). The fourth pair case 
($d=20.0\arcsec$) is
less obvious, since no clear signs of association are visible. Nevertheless
this object is listed as triple source in the ATESP 1.4 catalogue. 
The final 5~GHz
source list consists then in 111 distinct objects: 107 single sources and 4 
double sources.

The position of multiple sources is defined as the radio 
centroid, i.e. as the flux-weighted average position of all the components.
Integrated global source flux densities are computed by summing 
all the component integrated fluxes. The global source angular size is defined 
as \emph{largest angular size (las)} and it 
is computed as the maximum distance between the source components. 

We have also four sources which could not be parametrised by a single 
or multiple Gaussian fit.
{\it Non-Gaussian} sources have been parameterized as follows:
positions and peak flux densities have been derived by a second-degree 
interpolation of the flux density distribution. This means that positions 
refer to peak positions, which, for non-Gaussian sources does not
necessarily correspond to the position of the core. 
Integrated fluxes have been 
derived directly by summing pixel per pixel the flux density in the source 
area, defined as the region enclosed by the $\geq 3\sigma$ flux density 
contour. The source position angle was determined by the direction in which
the source is most extended and the source axes were defined again as 
{\it las}, i.e., in this case, the maximum distance between two opposite 
points belonging to the $3\sigma$ flux density contour along (major axis) and 
perpendicular to (minor axis) the same direction. 
These four non-Gaussian sources were catalogued 
at 1.4 GHz as follows: 2 non-Gaussian sources, 1 double source, 1 single 
Gaussian source. 

\subsection{Deconvolution}\label{sec-deconv}

The ratio of the integrated flux to the peak flux is a direct measure of the 
extension of a radio source:
\begin{equation}\label{eq-dec1}
S_{\rm total}/S_{\rm peak}=\theta_{\rm min} \; \theta_{\rm maj} 
/b_{\rm min} \; b_{\rm maj} 
\end{equation}
where $\theta_{\rm min}$ and $\theta_{\rm maj}$ are the source FWHM axes and 
$b_{\rm min}$ and $b_{\rm maj}$ are the synthesized beam FWHM axes. The
flux ratio can therefore be used to discriminate between extended (larger 
than the beam) and point-like sources. 

\begin{figure}[t]
\resizebox{\hsize}{!}{\includegraphics{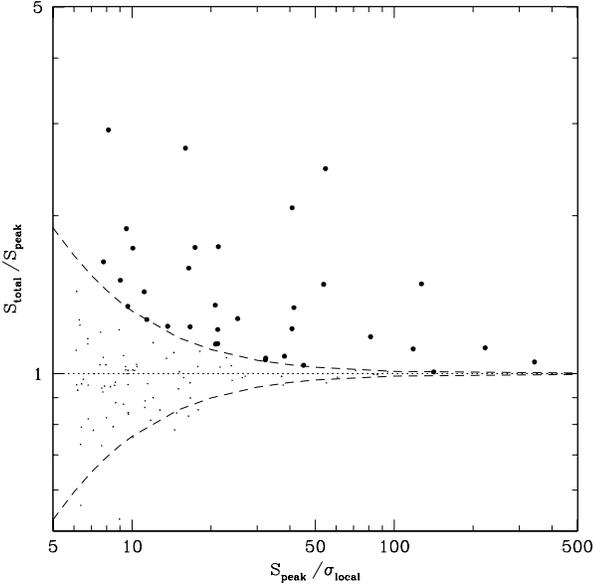}}
\caption{Ratio of the integrated flux $S_{\rm total}$ to the peak one 
$S_{\rm peak}$ as a function of the source signal-to-noise. The horizontal
dotted line indicates the $S_{\rm total}=S_{\rm peak}$ locus. 
Also shown (dashed lines) are the lower and upper 
envelopes (ref. Eq.~\ref{eq-dec2}) of the 
flux ratio distribution containing $\sim 90\%$ of the 
unresolved sources (dots). Filled circles indicate extended sources.
\label{fig-stspratio}}
\end{figure}

In Fig.~\ref{fig-stspratio} we show the flux 
ratio as a function of signal-to-noise for the 115 sources (or 
source components) detected above the $6\sigma_{\rm local}$--threshold. 
Values for 
$S_{\rm total}/S_{\rm peak} < 1$ are due to the influence of the image noise 
on the measure of source sizes and therefore of the source integrated fluxes.
Following Prandoni et al.~(\cite{Prandoni00b}, Sect.~3.1), we have taken into
account such errors by considering as unresolved all sources which lie below 
the curve defined by:
\begin{equation}\label{eq-dec2}
S_{\rm total}/S_{\rm peak} < 1 + 
\left[ \frac{ a }{ (S_{\rm peak}/\sigma_{\rm local})^{\beta}}\right]  
\end{equation}
where $a=10$ and $\beta=1.5$ (upper dashed line in 
Fig.~\ref{fig-stspratio}). Such curve is obtained by determining the lower 
envelope of the flux ratio distribution (the curve containing 90\% of the
$S_{\rm total}<S_{\rm peak}$ sources, see lower dashed line in 
Fig.~\ref{fig-stspratio}) and then mirroring it on the 
$S_{\rm total}>S_{\rm peak}$ side.

From this analysis we found that $\sim 70\%$ of the sources (or source 
components) have to be considered unresolved 
(dots in Fig.~\ref{fig-stspratio}).  
Deconvolved angular sizes are considered meaningful and given in the catalogue 
only for extended sources (filled circles in Fig.~\ref{fig-stspratio}); 
for unresolved sources, angular sizes are set to zero in the catalogue
(see Table~\ref{tab-5ghzlhrcat}).

\subsection{Source Parameters at full resolution}\label{sec-hrpar}

The sources extracted from the low resolution
mosaics and catalogued in Table~\ref{tab-5ghzlhrcat} were then searched for
in the full resolution mosaics, in order to get additional information on their
radio morphology. The detection threshold was set to $3\times$ the local noise 
level, as allowed whenever the source position is known a priori. 
The local noise was evaluated in a $\sim 1\arcmin \times 1\arcmin$ box 
centered on the source position. Above $3\sigma_{\rm local}$ we 
detected 109 of the 111 sources catalogued at low resolution. 

The source
full-resolution parameterization has been performed using similar procedures
as used at low resolution. Whenever the source peak flux densities were 
$\geq 6\sigma_{\rm local}$ we performed a two-dimensional Gaussian fit. 
The sources were then visually inspected and re-fitted, whenever necessary. 
We also applied Eq.~\ref{eq-dec2} (with $a=5$, $\beta=1.5$) 
to separate point sources from extended ones. 

The full-resolution 
parameters for each source are reported in Table~\ref{tab-5ghzlhrcat} 
(Columns 11-20). For sources detected at $<6\sigma_{\rm local}$ 
(flagged '$D$' in the catalogue) we 
provide only position and peak flux density. In two cases, where the
source is very extended and the structure is well recognized, we list all the 
parameters. Such cases are flagged '$DE$'. For the
two undetected sources (flagged '$U$' in the catalogue) we provide peak flux 
upper limits only. 

\begin{figure*}[t]

\vspace{-1.8cm}

\resizebox{\hsize}{!}{\includegraphics{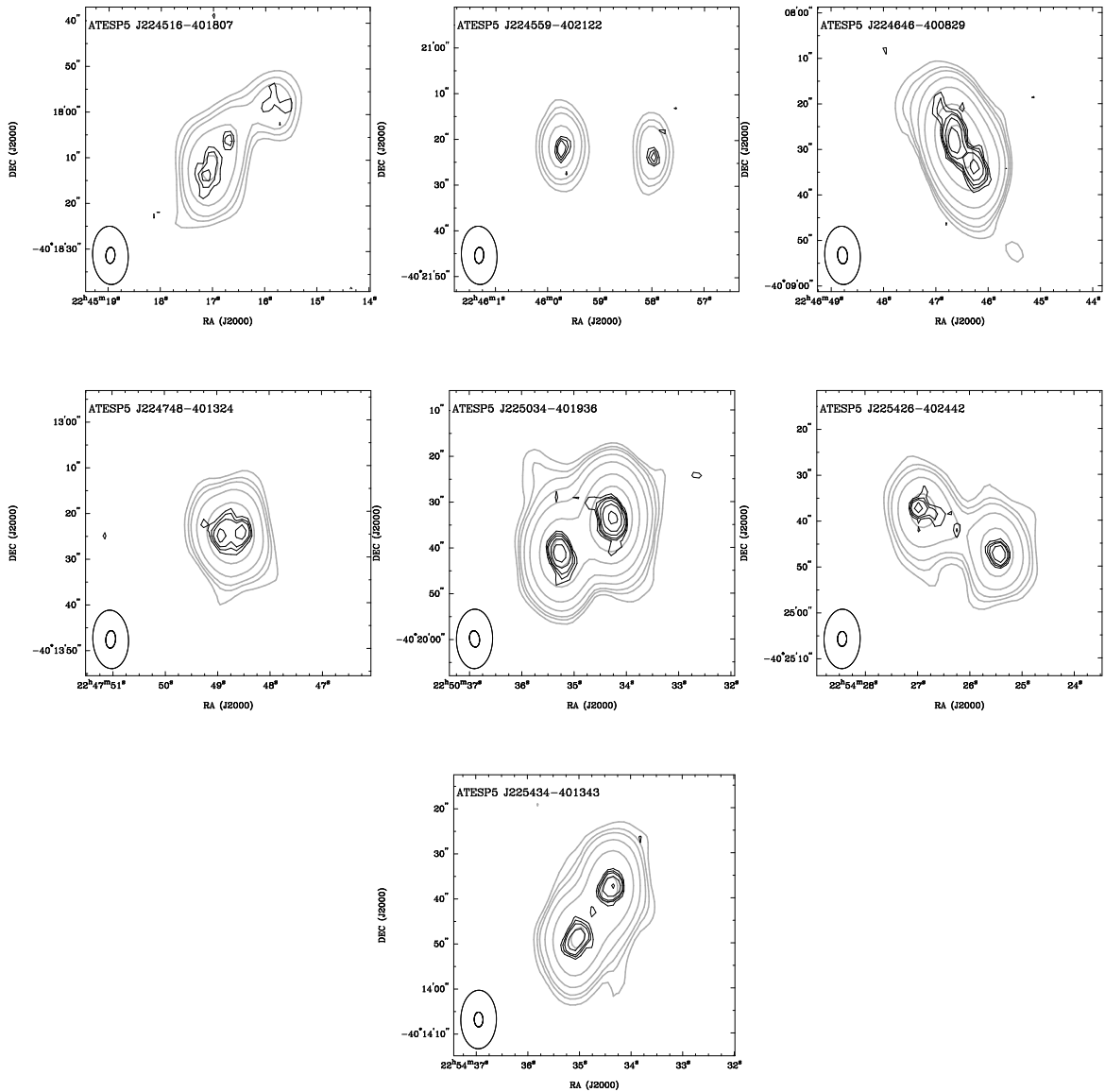}}

\vspace{-7cm}

\caption{$1\arcmin \times 1\arcmin$ contour images of multiple sources in the 
catalogue: 7 sources split in two or more components at either low or full 
resolution. Gray lines are for low resolution contours and black 
lines are for 
full resolution contours. For each source the contour levels are at 
3, 4.5, 6, 10, 20, 50, 100 $\times$ the local noise level. Synthesized beams 
at low and full resolution are shown in the lower left corner. 
\label{fig-multiple}}
\end{figure*}

We notice that 28 sources that are 
point-like at low resolution get deconvolved at full resolution, 
implying physical sizes of $1-3$ arcsec. In 3 additional cases, sources 
which appear as single at low resolution are split in two components at 
full resolution. Contour images of these three sources, together with the four 
sources catalogued as double at low resolution, are shown in 
Fig.~\ref{fig-multiple}. 

\subsection{Errors in the Source Parameters}\label{sec-errors}

Parameter uncertainties are the quadratic sum of two 
independent terms: the calibration errors, which usually dominate at high 
signal-to-noise ratios, and the internal errors, due to the presence of noise 
in the maps, which dominate at low signal-to-noise ratios. 

For an estimation of the internal errors in the 5~Ghz source parameters we 
refer to Condon's master equations (Condon~\cite{Condon97}), which provide
error estimates for elliptical Gaussian fitting procedures. 
Such equations already proved to be adequate 
to describe the measured internal errors for the ATESP 1.4~GHz source 
parameters (Prandoni et al. 2000b), which have been obtained 
in a very similar way: same detection algorithm (IMSAD) applied to very similar
radio mosaics.  

Flux calibration errors are in general estimated from comparison with 
consistent external data of better accuracy than the one tested. Due to the 
lack of  
data of this kind in the region covered by the ATESP 5~GHz survey, we assume 
the same calibration errors as for the ATESP 1.4 GHz survey, e.g. $5-10\%$ for
both flux densities and source sizes (see Appendix~A in 
Prandoni et al. 2000b). 

\section{The ATESP 5~GHz Sources}\label{sec-5ghzsources}

\subsection{The 5~GHz Source Catalogue}\label{sec-cat}

The 5 GHz source catalogue is reported in Table~\ref{tab-5ghzlhrcat}. 
The catalogue is sorted on right ascension. Each source is identified by
an IAU name (Column 1) and is defined by its low resolution parameters
(Columns $2 - 10$). 

The corresponding full resolution parameters are reported in
Columns $11 - 19$. 
The detailed format is the following:

{\it Column (1) -} Source IAU name. For multiple sources we list all 
the components (labeled `A', `B', etc.) preceded by a line (flagged {\it `M'}, 
see Column 9) giving the position of the radio 
centroid, the source global flux density and its overall angular size.

{\it Column (2) and (3) -} Source position: Right Ascension and 
Declination (J2000). 

{\it Column (4) and (5) -} Source peak ($S_{\rm peak}$) and 
integrated ($S_{\rm total}$) flux densities in mJy (Baars
et al. \cite{Baarsetal77} scale). 

{\it Column (6) and (7) -} Intrinsic (deconvolved from the beam) 
source angular size. Full width half maximum
of the major ($\Theta_{\rm maj}$) and minor ($\Theta_{\rm min}$) axes in 
arcsec. Zero values refer to unresolved sources (see Sect.~\ref{sec-deconv}).

{\it Column (8) -} Source position angle (P.A., measured N through E) 
for the major axis in degrees. 

{\it Column (9) -} Flag indicating the fitting procedure and 
parameterization adopted for the source (or source component). 
{\it `S'} refers to Gaussian fits; {\it `S*'} refers to poor Gaussian fits 
(see Sect.~\ref{sec-det}); {\it `E'} refers to non-Gaussian sources and 
{\it `M'} refers to multiple sources (see Sect.~\ref{sec-mult}). 

{\it Column (10) -} Local noise level used for the source detection in $\mu$Jy
(see Sects.~\ref{sec-mosanalysis} and ~\ref{sec-det}).  

{\it Columns (11) to (17) -} Same as Columns (2) to (8), but 
referring to full resolution source parameterization.

{\it Column (18) -} Same as Column (9), but 
referring to full resolution source parameterization. Here additional 
flags are defined for any sources with peak flux  
$3\sigma_{\rm local}\leq S_{\rm{peak}} < 6\sigma_{\rm local}$ and undetected 
sources (see Sect.\ref{sec-hrpar}). 
In a few cases references are given to additional notes on full resolution 
source radio morphology (reported as footnotes at the end of the table). 

{\it Column (19) -} Same as Column (10) but referring to full resolution 
mosaics (see Sect.~\ref{sec-hrpar}).  

\begin{figure}[t]
\resizebox{\hsize}{!}{\includegraphics{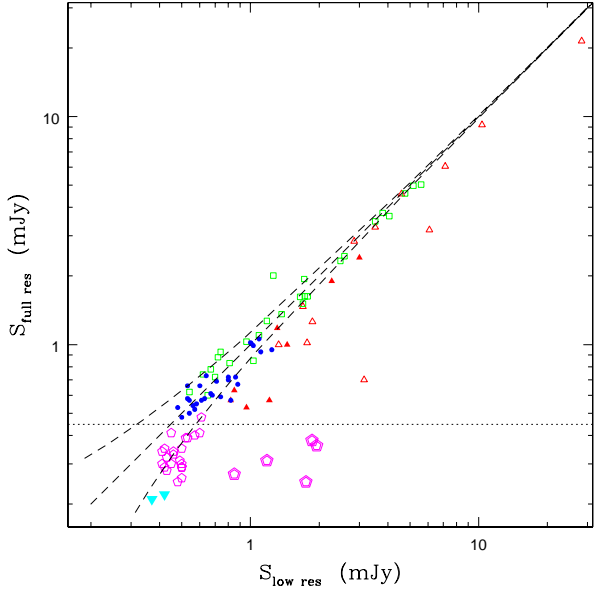}}
\caption{Low resolution vs. full resolution flux
densities for the ATESP 5~GHz sample (multiple-component sources are 
not plotted). We define the flux as $S_{\rm{total}}$ whenever a source is 
resolved and as $S_{\rm{peak}}$ otherwise.
Dashed lines represent the $S_{\rm{full \, res.}}=S_{\rm{low \, res}}$ line 
and the $1\sigma$ confidence limits in the flux measurements. 
The horizontal dotted 
line indicates the average $6\sigma$ detection limit at full resolution. 
Blue dots: sources appearing point-like at both low and full resolution; 
green empty squares: sources that are point-like at low resolution, but 
resolved at full resolution; 
magenta pentagons: sources point-like at low resolution and 
detected at $<6\sigma_{\rm{local}}$ at full resolution (flagged '$D$' in 
Column~18 of Table~\ref{tab-5ghzlhrcat}); 
cyan filled triangles (pointing downwards): sources that are 
point-like at low 
resolution but undetected at full resolution (flagged '$U$' in Column~18 of 
Table~\ref{tab-5ghzlhrcat}); 
red empty triangles: sources resolved at both low and full resolution;
red filled triangles: sources resolved at low resolution and point-like at 
full resolution; bold magenta pentagons: sources resolved at low 
resolution and detected at $<6\sigma_{\rm{local}}$ at full resolution 
(flagged '$D$' or '$DE$'
in Column~18 of Table~\ref{tab-5ghzlhrcat}).    
\label{fig-fluxlrhr}}
\end{figure}

\begin{figure}[t]
\resizebox{\hsize}{!}{\includegraphics{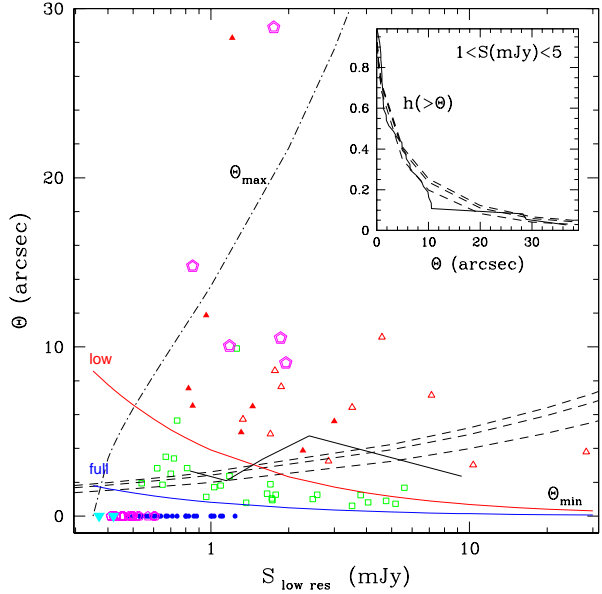}}

\caption{Deconvolved angular size ($\Theta = \Theta_{\rm{maj}}$)
as a function of integrated flux density (same sources and symbols as in 
Fig.~\ref{fig-fluxlrhr}). We plot low resolution source parameters,  
except when the source is unresolved at low resolution: 
in that case we use the
full resolution $\Theta_{\rm{maj}}$ estimation.
For sources unresolved at both low and full resolution 
we set $\Theta=0$ and use the peak flux density. 
The dot-dashed line represents the size ($\Theta_{\rm max}$) above which
the ATESP 5 GHz sample gets uncomplete, due to the resolution bias.
The two solid lines indicates the minimum angular size ($\Theta_{\rm min}$), 
below 
which deconvolution is not considered meaningful at low (red upper line) 
and at full (blue lower line) resolution respectively. 
Dashed lines indicate the median source sizes as a function of flux as 
expected from the Windhorst et al. (1990) relation. 
Such lines have to be compared to the broken solid
line which represents the median source sizes for different flux intervals, as 
measured in our sample. The inner panel shows the angular size 
distribution ($h(>\Theta)$ proposed by Windhorst et al. (dashed lines),
compared to the size distribution of sources in our sample (broken solid line).
See text for more details.
\label{fig-sizeflux}}
\end{figure}

\subsection{Source Structure and Size}\label{sec-sizes}

A comparison between full and low resolution source parameters gives 
interesting information about the source size and structure. 

\begin{figure*}[t]

\vspace{-2.cm}

\resizebox{\hsize}{!}{\includegraphics{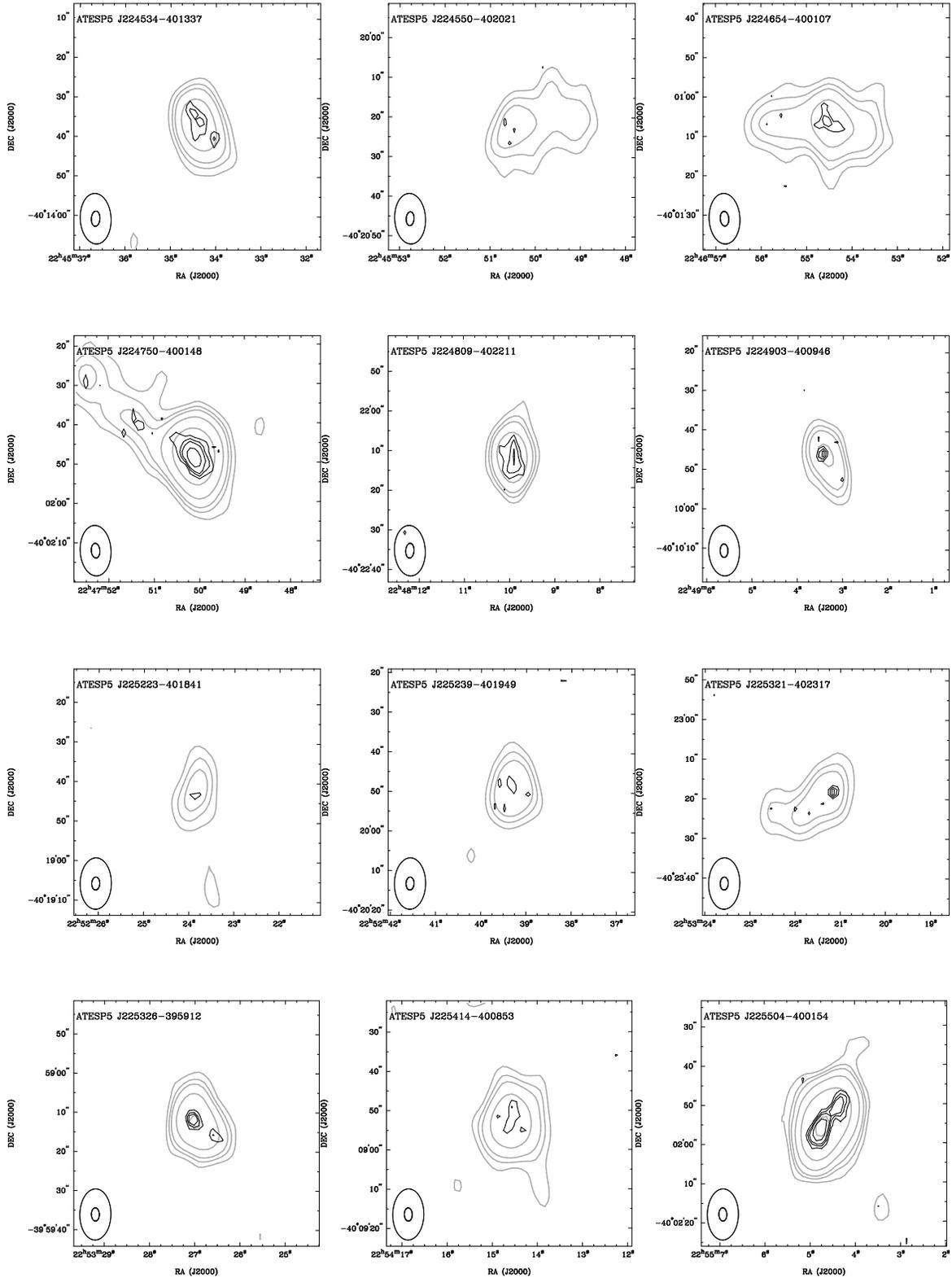}}

\vspace{-1.5cm}

\caption{$1\arcmin \times 1\arcmin$ contour images of most extended sources in 
the catalogue: 12 sources with deconvolved size (at either low or full 
resolution) larger than $8\arcsec$. Gray lines are for low 
resolution contours and black lines are for full resolution contours. 
For each source the contour levels are at 3, 4.5, 6, 10, 20, 50, 100 $\times$ 
the local noise level. Synthesized beams 
at low and full resolution are shown in the lower left corner. 
\label{fig-extended}}
\end{figure*}

In Fig.~\ref{fig-fluxlrhr} we compare the flux densities at low and full 
resolution for the 104 ATESP 5~GHz sources, catalogued as single sources
(we exclude the 7 multi--component sources shown in Fig.~\ref{fig-multiple}). 
We find a good correlation between the low and full resolution fluxes,  
although a number of sources show a systematic integrated flux underestimation 
at the higher spatial resolution. 
Such a resolution effect is due to the loss of low surface brightness flux
in extended sources, as it is evident from the comparison between 
Fig.~\ref{fig-fluxlrhr} and Fig.~\ref{fig-sizeflux}, where we plot the 
source deconvolved angular sizes as a function of flux density. 
In fact the largest sources are typically the ones with largest
flux underestimations at full resolution.  
The most extreme cases regard sources which are 
resolved at low resolution and are catalogued as point-like (red filled 
triangles) or even as $<6\sigma_{\rm{local}}$ detections (bold magenta 
pentagons) at full resolution. In such cases the extended flux
gets almost entirely resolved out. We illustrate this effect 
in Fig.~\ref{fig-extended}, where we show contour images of the 12 most 
extended 
single--component sources ($\Theta_{\rm maj} > 8\arcsec$). Among such
sources there are several cases of low surface brightness sources, which are 
barely detected at full resolution, and a few cases where full resolution 
reveals elongated structures and/or hints of multiple components. 

The two solid lines in Fig.~\ref{fig-sizeflux} indicate the minimum angular 
size, $\Theta_{\rm min}$, below which deconvolution is not considered 
meaningful at either low (red upper line) or full resolution (blue lower line).
They are derived from Eqs.~(\ref{eq-dec1}) and (\ref{eq-dec2}), by setting the
appropriate $b_{\rm min}$, $b_{\rm maj}$, $a$ and $\beta$ parameters. 
We notice that full resolution allows us to get size information for sources
with intrinsic sizes ranging from $1-2$ arcsec to $\sim 4$ arcsec (green empty 
squares).
This means that we can successfully deconvolve $\sim 80\%$ of the ATESP 5~GHz
sources with $S_{\rm low \, res} > 0.7$ mJy and $\sim 90\%$ of the sources 
with $S_{\rm low \, res}>1$ mJy. 
Above such flux limits, where we have a limited number of upper limits, 
we can reliably undertake a statistical analysis of the size properties of our 
sample. In Fig.~\ref{fig-sizeflux} we compare the median angular size measured 
in different flux intervals for the ATESP 5~GHz sources with $S>0.7$ mJy 
(broken solid line) and the angular size integral distribution derived for the 
ATESP 5~GHz sources with $1<S(\rm{mJy})<5$ (broken solid line in the inner 
panel) to the ones obtained from the Windhorst et al. (1990) relations
proposed for 1.4 GHz samples:
$\Theta_{\rm med} = 2\arcsec \cdot S_{1.4 \; \rm GHz}^{0.30}$ ($S$ in mJy) and 
$h(>\Theta)=e^{-ln{2}\;(\Theta/\Theta_{\rm med})^{0.62}}$.
$\Theta_{\rm med}$ and $h(>\Theta)$ are extrapolated to 5~GHz using three 
different values for the spectral index: $\alpha = 0$, -0.5 and -0.7 (see 
black dashed lines). We notice that in the flux ranges considered our 
determinations show a very good agreement with the ones of Windhorst et al. 
(1990). 

On the other hand, we are not able to say much 
about the faintest sources 
($S<0.7$ mJy), which remain largely unresolved even at full resolution, 
and constitute about 40\% of our sample.
We can only argue that the angular sizes of such sources are smaller 
than $1-2$ arcsec. Such upper limits are consistent with the size analysis 
performed by Fomalont et al. (1991) for a deeper sample selected at the same 
frequency (5~GHz): $16/19$ ($84\%$) sources with $0.06<S(\rm{mJy})<0.8$ 
have $\Theta < 1.5$ arcsec. On the other hand, the median values derived 
from the Windhorst et al. relation (transformed to 5 GHz) might be possibly
overestimating the real values (compare black dashed lines to full resolution
blue solid line in Fig.~\ref{fig-sizeflux}).  

We finally notice that flux losses in extended sources not only affect the 
full resolution parameterization of the ATESP 5~GHz sources 
(as shown in Fig.~\ref{fig-fluxlrhr}), but can also cause incompletess in the 
ATESP 5~GHz catalogue itself (defined at low resolution). A resolved source 
of given $S_{\rm total}$ will drop below the peak flux density 
detection threshold more easily than a point source of same $S_{\rm total}$. 
This is the so--called {\it resolution bias}. 

Eq.~(\ref{eq-dec1}) -- with the low resolution parameter setting -- 
can be used to give an approximate estimate of the maximum size 
($\Theta_{\rm max}$) a source of given $S_{\rm total}$ can have 
before dropping below the $S_{\rm peak}=6\sigma_{\rm local}$ limit of the 
ATESP 5~GHz catalogue. Such limit is represented by the black dot-dashed line 
plotted in Fig.~\ref{fig-sizeflux}.
As expected, the angular sizes of the largest ATESP 5 GHz sources  
approximately follow the estimated $\Theta_{\rm max}-S$ relation.

In principle there is a second incompleteness effect, related to
the maximum scale at which the ATESP 5~GHz low resolution mosaics are 
sensitive due to the lack of baselines shorter than 60 m. According to 
it, we expect the ATESP 5~GHz sample to become progressively insensitive to 
sources larger than $90\arcsec$ (see Sect.~\ref{sec-mosprod}).
This latter effect can, however, be neglected in this case, 
because it is smaller than the previous one over the entire flux 
range spanned by the ATESP 5 GHz survey. Moreover, if we assume the angular 
size distribution proposed by Windhorst et al. (1990), we expect 
$<1$ sources with $\Theta>90\arcsec$ in the area and flux range covered by our
survey. 

\section{Results}\label{sec-results}

\subsection{The ATESP 5~GHz Source Counts}\label{sec-counts}

\begin{figure*}[t]
\resizebox{12cm}{!}{\includegraphics{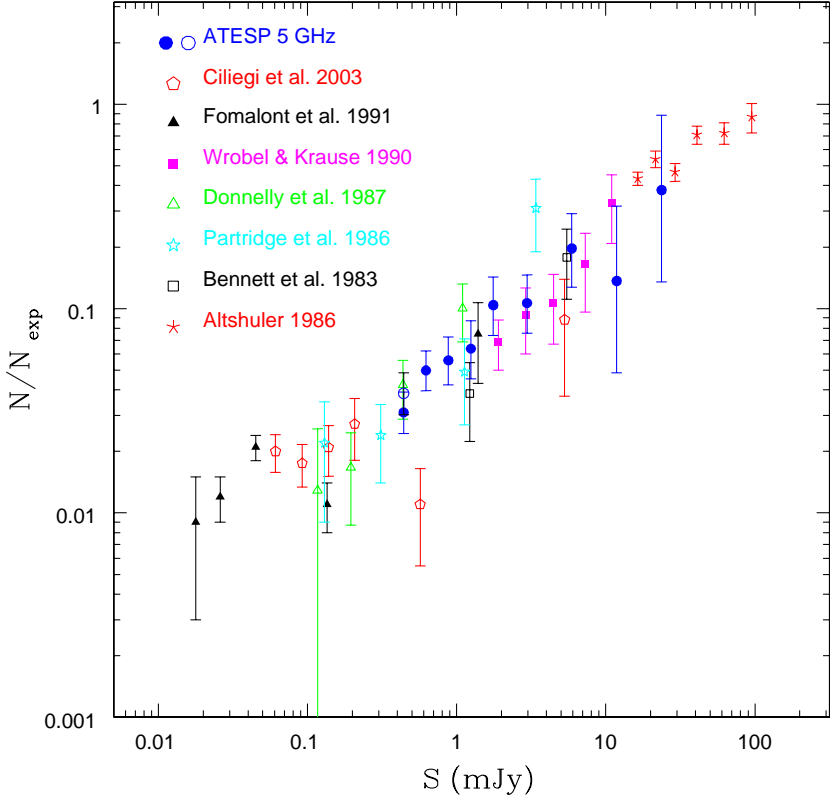}}
\hfill
\parbox[b]{55mm}{
\caption{Normalized 5 GHz differential source counts for 
different samples: Ciliegi et al. 2003 (empty pentagons), Fomalont et al. 1991 
(filled triangles), Wrobel \& Krause 1990 (filled squares), Donnelly et al. 
1987 (empty triangles), Partridge et al. 1986 (stars), Altshuler 1986 
(asterisks), Bennett et al. 1983 (empty squares).
The ATESP 5 GHz source counts presented here (filled circles plus single empty 
circle) are corrected for resolution bias as explained in the text. 
Vertical bars represent Poissonian errors on the normalized counts.}
\label{fig-counts}}
\end{figure*}

\begin{table}[t]
\caption{ATESP 5 GHz source counts. 
\label{tab-diffcounts}}

\begin{flushleft}
\begin{tabular}{ccrcc}
\hline\hline\noalign{\smallskip}
\multicolumn{1}{c}{$\Delta S$}
& \multicolumn{1}{c}{$<S>$}
&\multicolumn{1}{c}{$N_S$}
&\multicolumn{1}{c}{$dN/dS$}
& \multicolumn{1}{c}{$N/N_{\rm exp}$}\\
\multicolumn{1}{c}{(mJy)}
& \multicolumn{1}{c}{(mJy)}
&\multicolumn{1}{c}{}
&\multicolumn{1}{c}{sr$^{-1}$ Jy$^{-1}$}
& \multicolumn{1}{c}{}\\
\noalign{\smallskip}
\hline\noalign{\smallskip} 
0.37 -- 0.52 &  0.44 & 22  &  $6.92\times 10^8$ & 0.031$_{-0.007}^{+0.008}$ \\
\noalign{\medskip}
0.52 -- 0.74 &  0.62 & 24  & $4.67\times 10^8$ &  0.050$_{-0.010}^{+0.010}$ \\
\noalign{\medskip}
0.74 -- 1.05 &   0.88  & 17 & $2.20\times 10^8$ & 0.056$_{-0.014}^{+0.017}$ \\
\noalign{\medskip}
1.05 -- 1.48 &  1.24  & 12  & $1.05\times 10^8$ & 0.064$_{-0.018}^{+0.024}$ \\
\noalign{\medskip}
1.48 -- 2.09 &   1.76 & 12 & $7.24\times 10^7$ &  0.104$_{-0.030}^{+0.040}$ \\
\noalign{\medskip}
2.09 -- 4.19 &   2.96 & 12 & $2.06\times 10^7$ &  0.106$_{-0.031}^{+0.040}$ \\
\noalign{\medskip}
4.19 -- 8.37 &   5.92  & 8 & $6.74\times 10^6$ &  0.197$_{-0.070}^{+0.094}$ \\
\noalign{\medskip}
8.37 -- 16.7 &  11.8 & 2 &  $8.26\times 10^5$ &  0.137$_{-0.088}^{+0.180}$ \\
\noalign{\medskip}
16.7 -- 33.5 &  23.7 & 2 &  $4.06\times 10^5$ &  $0.380_{-0.245}^{+0.502}$ \\
\noalign{\medskip}
\hline
\end{tabular}
\end{flushleft}
\end{table}

We used the 111 5 GHz ATESP sources to derive the differential source 
counts as a function of flux density. 
In computing the counts we have used the catalogue as defined at low 
resolution to minimize incompleteness induced by resolution effects. 
Integrated flux densities were used for extended sources and 
peak flux densities for point-like sources. 

Each source has been weighted for the reciprocal of its visibility area 
($\Omega(S_{\rm peak})/\Omega_{\rm tot}$, see Fig.~\ref{fig-areaflux}), 
that is the fraction of the total area over which 
the source could be detected. We
notice that for $S_{\rm peak}>0.5$ mJy a source can be counted over the whole 
survey area. Moreover, we have taken properly into account the catalogue 
incompleteness in terms of integrated flux density, due to the resolution 
bias discussed in the previous section. 
The correction $c$ for the resolution bias has been defined following 
Prandoni et al. (2001a) as: 
\begin{equation}\label{eq-rescorr}
c=\frac{1}{1-h(>\Theta_{\rm lim})}
\end{equation}
where $h(>\Theta_{\rm lim})$ is the integral angular size distribution 
proposed by Windhorst et al. (1990) for 1.4 GHz samples, which turned out 
to be a good representation of the ATESP 5~GHz source sizes at least down 
to $S\sim 0.7$ mJy (see Sect.~\ref{sec-sizes}). Here we transform the 
Windhorst et al. relation to 5 GHz using $\alpha=-0.5$, chosen as reference 
value (see also Sect.~\ref{sec-spindex}). 

$\Theta_{\rm lim}$ represents the angular size upper limit, above which
we expect to be incomplete. This is defined as a function of the integrated 
source flux density as (see Prandoni et al. 2001a):
\begin{equation}\label{eq-thetalim}
\Theta_{\rm lim} = \mbox{max} [ \Theta_{\rm min},\Theta_{\rm max} ]
\end{equation}
where $\Theta_{\rm min}$ and $\Theta_{\rm max}$ are the parameters defined
in Sect.~\ref{sec-sizes}. The $\Theta_{\rm min}-S$ relation (red upper solid 
line in Fig.~\ref{fig-sizeflux}) is 
important at low flux levels where $\Theta_{\rm max}$ (black dot-dashed 
line in Fig.~\ref{fig-sizeflux}) becomes unphysical 
(i.e. $\rightarrow 0$). In other words, introducing $\Theta_{\rm min}$ in the 
equation takes into account the effect of having a finite synthesized beam size
(that is $\Theta_{\rm lim}>>0$ at the survey limit) and a deconvolution
efficiency which varies with the source peak flux. 

The 5 GHz ATESP source counts are shown in Fig.~\ref{fig-counts} (filled 
circles) and listed in Table~\ref{tab-diffcounts}, where,
for each flux interval ($\Delta S$), the geometric mean of the flux 
density ($<S>$), 
the number of sources detected ($N_S$), the differential source density 
($dN/dS$) and the normalized
differential counts ($N/N_{\rm exp}$) are given. Also listed 
are the Poissonian errors (calculated following Regener 1951) associated to 
the normalized counts. 
For comparison with other 5~GHz studies, the source counts are normalized to
a non evolving Euclidean model which fits the brightest sources in the sky.
We notice that at 5 GHz the standard Euclidean integral counts are 
$N(>S)=60\times S^{-1.5}$ sr$^{-1}$ ($S$ in Jy).

In determining the resolution bias incompleteness, correct definitions of  
$h(>\Theta_{\rm lim})$ and $\Theta_{\rm lim}$ could be very important. 
We checked the robustness of the resolution bias correction by setting 
different values for $\alpha$ when transforming $h(>\Theta_{\rm lim})$ to 
5 GHz and we changed the definition of $\Theta_{\rm lim}$. 
In particular we tested the effect of setting a fixed value for 
$\Theta_{\rm min}$ in Eq.~(\ref{eq-thetalim}). In this case 
$\Theta_{\rm min}$ was set equal to $4\arcsec$, which roughly represents the 
typical minimum angular size that is reliably deconvolved at low resolution 
(see Fig.~\ref{fig-sizeflux}). This latter definition has the effect of
increasing the resolution bias correction at the faintest fluxes. The same 
happens when decreasing the spectral index from 0 to -0.7. In general, 
however, significant changes in the normalized counts are seen only for 
the faintest flux bin. To illustrate such effect at low fluxes, in 
Fig.~\ref{fig-counts} we add the point obtained by computing the resolution 
bias correction $c$ with $\Theta_{\rm min}=4\arcsec$ and $\alpha=-0.7$ 
(empty circle at $S=0.44$ mJy). 

The ATESP source counts are compared with previous 
determinations at 5 GHz. As illustrated in Fig.~\ref{fig-counts} 
the ATESP counts are consistent with others reported in the 
literature and improve significantly the statistics available
in the flux range $0.4 - 1$ mJy. The ATESP 5~GHz counts 
do not show evidence of flattening or slope change down to the survey 
limit. This fact is discussed in view of the spectral index properties of 
the ATESP radio sources in Sect.~\ref{sec-spindex}. 

\subsection{$1.4 - 5$ GHz spectral index analysis}\label{sec-spindex}

\begin{table*}[t]
\caption{$1.4-5$ GHz spectral index of ATESP sources. 
\label{tab-si}}
\scriptsize
\begin{flushleft}
\begin{tabular}{ccrrr|ccrrr}
\hline\hline\noalign{\smallskip}
\multicolumn{1}{c}{5~GHz IAU name}
& \multicolumn{1}{c}{1.4~GHz IAU name}
& \multicolumn{1}{c}{$S_{\rm 5 \, GHz}$}
&\multicolumn{1}{c}{$S_{\rm 1.4 \, GHz}$}
&\multicolumn{1}{r|}{$\alpha \; \pm \; \sigma(\alpha)$} &
\multicolumn{1}{c}{5~GHz IAU name}
& \multicolumn{1}{c}{1.4~GHz IAU name}
& \multicolumn{1}{c}{$S_{\rm 5 \, GHz}$}
&\multicolumn{1}{c}{$S_{\rm 1.4 \, GHz}$}
&\multicolumn{1}{r}{$\alpha \; \pm \; \sigma(\alpha)$}\\
\multicolumn{1}{c}{(ATESP5 J...)} & \multicolumn{1}{c}{(ATESP J...)} 
& \multicolumn{1}{c}{(mJy)}& \multicolumn{1}{c}{(mJy)}
& \multicolumn{1}{c|}{} &
\multicolumn{1}{c}{(ATESP5 J...)} & \multicolumn{1}{c}{(ATESP J...)} 
& \multicolumn{1}{c}{(mJy)}& \multicolumn{1}{c}{(mJy)}
& \multicolumn{1}{c}{}\\
\noalign{\smallskip}
\hline\noalign{\smallskip} 
&&&&&&&& \\
 J224502-400415 &    J224502-400415 &  0.65  &  1.12  &  $-0.43 \pm 0.21 $ &	- &  J224911-400859 &  0.36  &  0.88  &  $-0.70 \pm 0.33 $ \\
 J224509-400622 &    J224509-400623 &  0.70  &  2.45  &  $-0.98 \pm 0.16 $ &	- &  J224917-401330 &  $<$0.19  &  0.61  &  $<-0.92 $ \\
 J224510-401655 &    J224510-401657 &  0.46  &  0.57  &  $-0.17 \pm 0.36 $ &	 J224919-400037 &    J224919-400037 &  0.64  &  0.91  &  $-0.28 \pm 0.23 $ \\
 J224513-400052 &    J224513-400051 &  0.46  &  0.59  &  $-0.20 \pm 0.34 $ &	 J224932-395801 &    J224932-395800 &  0.45  &  0.55  &  $-0.16 \pm 0.36 $ \\
- &  J224513-400407 &  0.27  &  0.58  &  $-0.60 \pm 0.48 $ &	 J224935-400816 &    J224935-400816 &  0.82  &  0.70  &  $0.12 \pm 0.23 $ \\
 J224516-401807 &    J224516-401807 &  3.87  &  9.54  &  $-0.71 \pm 0.04 $ &	 J224948-395918 &    J224948-395920 &  1.72  &  0.87  &  $0.54 \pm 0.17 $ \\
 J224518-401001 &    J224518-401001 &  5.60  &  10.37  &  $-0.48 \pm 0.02 $ &	 J224951-402035 &    J224951-402233 &  0.50  &  1.62  &  $-0.92 \pm 0.24 $ \\
 J224530-401141 &   - &  0.88  &  0.49  &  $0.46 \pm 0.30 $ &	 J224958-395855 &    J224958-395855 &  1.65  &  1.52  &  $0.06 \pm 0.12 $ \\
 J224533-402014 &    J224533-402015 &  0.53  &  1.04  &  $-0.53 \pm 0.27 $ &	 J225004-402412 &    J225004-402413 &  1.78  &  3.16  &  $-0.45 \pm 0.08 $ \\
 J224534-401337 &    J224534-401337 &  1.86  &  3.79  &  $-0.56 \pm 0.07 $ &	 J225008-400425 &    J225008-400425 &  1.70  &  2.88  &  $-0.41 \pm 0.09 $ \\
 J224534-400049 &    J224534-400050 &  1.87  &  6.02  &  $-0.92 \pm 0.07 $ &	- &  J225009-400605 &  0.40  &  0.84  &  $-0.58 \pm 0.34 $ \\
- &  J224535-402531 &  $<$0.20  &  0.57  &  $<-0.82 $ &	 J225028-400333 &    J225029-400332 &  0.42  &  0.75  &  $-0.46 \pm 0.34 $ \\
 J224547-400324 &    J224547-400324 &  28.28  &  32.83  &  $-0.12 \pm 0.01 $ &	 J225034-401936 &    J225034-401936 &  25.78  &  76.62  &  $-0.856 \pm 0.004 $ \\
 J224550-402021 &    J224550-402022 &  1.75  &  3.54  &  $-0.55 \pm 0.08 $ &	 J225048-400147 &   - &  0.96  &  0.60  &  $0.37 \pm 0.27 $ \\
 J224551-401618 &    J224551-401619 &  0.57  &  1.16  &  $-0.56 \pm 0.24 $ &	 J225056-402254 &   - &  0.43  &  0.41  &  $0.04 \pm 0.45 $ \\
 J224557-400934 &    J224557-400931 &  0.42  &  1.35  &  $-0.92 \pm 0.28 $ &	 J225056-400033 &    J225056-400033 &  2.27  &  1.17  &  $0.52 \pm 0.14 $ \\
 J224559-402122 &    J224558-402122 &  1.72  &  5.59  &  $-0.93 \pm 0.08 $ &	 J225057-401522 &    J225057-401522 &  3.00  &  2.01  &  $0.31 \pm 0.08 $ \\
 J224601-401502 &    J224601-401502 &  2.58  &  6.67  &  $-0.75 \pm 0.05 $ &	 J225058-401645 &   - &  0.50  &  0.45  &  $0.08 \pm 0.41 $ \\
 J224608-400414 &    J224608-400415 &  0.61  &  0.59  &  $0.03 \pm 0.29 $ &	 J225100-400934 &    J225100-400933 &  0.49  &  1.65  &  $-0.95 \pm 0.24 $ \\
 J224613-401132 &    J224613-401132 &  1.24  &  2.23  &  $-0.46 \pm 0.11 $ &	 J225112-402230 &    J225112-402229 &  1.03  &  3.06  &  $-0.86 \pm 0.13 $ \\
 J224623-400854 &    J224623-400856 &  0.68  &  2.26  &  $-0.94 \pm 0.17 $ &	 J225118-402653 &    J225118-402652 &  2.85  &  3.69  &  $-0.20 \pm 0.06 $ \\
- &  J224623-401845 &  0.41  &  1.44  &  $-0.99 \pm 0.31 $ &	 J225122-402524 &    J225122-402524 &  0.54  &  1.56  &  $-0.83 \pm 0.24 $ \\
 J224628-401207 &    J224628-401207 &  0.82  &  1.35  &  $-0.39 \pm 0.17 $ &	 J225138-401747 &    J225138-401747 &  0.62  &  1.61  &  $-0.75 \pm 0.20 $ \\
 J224632-400319 &    J224632-400320 &  0.72  &  1.83  &  $-0.73 \pm 0.18 $ &	 J225154-401051 &    J225155-401050 &  1.31  &  1.52  &  $-0.12 \pm 0.12 $ \\
- &  J224640-401710 &  0.27  &  0.65  &  $-0.69 \pm 0.47 $ &	- &  J225206-401947 &  0.33  &  1.25  &  $-1.05 \pm 0.36 $ \\
 J224646-400829 &    J224646-400830 &  13.25  &  38.2  &  $-0.83 \pm 0.01 $ &	 J225207-400720 &    J225207-400720 &  4.06  &  8.99  &  $-0.62 \pm 0.03 $ \\
 J224647-401220 &    J224647-401220 &  0.81  &  1.17  &  $-0.29 \pm 0.18 $ &	 J225217-402135 &    J225217-402136 &  1.73  &  2.16  &  $-0.17 \pm 0.09 $ \\
- &  J224648-400250 &  $<$0.20  &  1.68  &  $<-1.67 $ &	 J225223-401841 &    J225223-401843 &  0.85  &  0.98  &  $-0.11 \pm 0.20 $ \\
 J224654-400107 &    J224654-400108 &  3.15  &  5.59  &  $-0.45 \pm 0.05 $ &	 J225224-402549 &    J225224-402549 &  4.76  &  7.19  &  $-0.32 \pm 0.03 $ \\
- &  J224658-401206 &  $<$0.22  &  0.54  &  $<-0.71 $ &	 J225239-401949 &    J225239-401948 &  1.18  &  2.26  &  $-0.51 \pm 0.12 $ \\
 J224701-402646 &    J224701-402645 &  1.33  &  2.76  &  $-0.57 \pm 0.11 $ &	 J225242-395949 &    J2252342-395949 &  0.53  &  1.48  &  $-0.81 \pm 0.22 $ \\
 J224702-400948 &    J224702-400946 &  0.60  &  0.61  &  $-0.01 \pm 0.29 $ &	 J225249-401256 &    J225249-401256 &  1.09  &  1.52  &  $-0.26 \pm 0.14 $ \\
 J224707-400616 &    J224707-400616 &  1.70  &  3.37  &  $-0.54 \pm 0.08 $ &	 J225316-401200 &    J225316-401200 &  0.50  &  0.78  &  $-0.35 \pm 0.28 $ \\
 J224714-401453 &    J224714-401454 &  3.52  &  6.26  &  $-0.45 \pm 0.04 $ &	 J225321-402317 &    J225321-402319 &  1.21  &  2.32  &  $-0.51 \pm 0.11 $ \\
 J224714-402400 &   - &  0.57  &  0.45  &  $0.19 \pm 0.36 $ &	 J225322-401931 &    J225322-401931 &  1.01  &  1.86  &  $-0.48 \pm 0.14 $ \\
 J224719-400141 &    J224719-400142 &  0.48  &  1.32  &  $-0.79 \pm 0.27 $ &	 J225323-400453 &   - &  0.85  &  0.51  &  $0.40 \pm 0.31 $ \\
 J224719-401530 &    J224719-401531 &  0.74  &  0.84  &  $-0.10 \pm 0.23 $ &	 J225325-400221 &   - &  0.53  &  0.56  &  $-0.04 \pm 0.32 $ \\
 J224721-402043 &   - &  0.50  &  0.37  &  $0.24 \pm 0.43 $ &	 J225326-395912 &    J225326-395911 &  1.77  &  3.72  &  $-0.58 \pm 0.07 $ \\
 J224724-395909 &    J224724-395909 &  1.45  &  1.18  &  $0.16 \pm 0.14 $ &	 J225332-402721 &   - &  1.00  &  $<$0.26  &  $>1.06  $ \\
 J224727-402751 &    J224727-402750 &  0.67  &  1.94  &  $-0.84 \pm 0.21 $ &	 J225334-401414 &   - &  0.54  &  0.45  &  $0.14 \pm 0.38 $ \\
 J224727-401926 &   - &  1.11  &  0.32  &  $0.98 \pm 0.42 $ &	 J225344-401928 &   - &  3.52  &  0.60  &  $1.39 \pm 0.25 $ \\
 J224729-402000 &    J224729-402001 &  0.5  &  0.84  &  $-0.41 \pm 0.29 $ &	 J225345-401845 &   - &  0.48  &  0.60  &  $-0.18 \pm 0.33 $ \\
 J224731-400527 &   - &  0.43  &  0.35  &  $0.16 \pm 0.46 $ &	- &  J225351-400441 &  $<$0.17  &  0.96  &  $<-1.36  $ \\
 J224732-401442 &    J224732-401442 &  7.14  &  20.28  &  $-0.82 \pm 0.02 $ &	 J225353-400154 &    J225353-400153 &  1.03  &  1.10  &  $-0.05 \pm 0.17 $ \\
 J224735-402321 &    J224735-402321 &  1.18  &  1.12  &  $0.04 \pm 0.16 $ &	- &  J225354-400241 &  0.36  &  1.53  &  $-1.14 \pm 0.31 $ \\
 J224740-401821 &    J224740-401821 &  0.74  &  2.15  &  $-0.84 \pm 0.18 $ &	 J225400-402204 &    J225400-402204 &  0.54  &  1.25  &  $-0.66 \pm 0.24 $ \\
 J224741-400442 &    J224741-400443 &  0.52  &  0.61  &  $-0.13 \pm 0.31 $ &	 J225404-402226 &    J225404-402226 &  3.80  &  10.34  &  $-0.79 \pm 0.03 $ \\
 J224748-401324 &    J224748-401324 &  3.97  &  14.52  &  $-1.02 \pm 0.03 $ &	 J225414-400853 &    J225414-400852 &  1.95  &  3.10  &  $-0.36 \pm 0.07 $ \\
 J224750-400148 &    J224750-400143 &  6.09  &  13.44  &  $-0.62 \pm 0.02 $ &	 J225426-402442 &    J225426-402442 &  4.24  &  10.43  &  $-0.71 \pm 0.03 $ \\
 J224753-400455 &    J224753-400456 &  0.67  &  2.08  &  $-0.89 \pm 0.19 $ &	 J225430-400334 &   - &  0.63  &  $<$0.26  &  $>0.70 $ \\
- &  J224759-400825 &  0.21  &  0.82  &  $-1.07 \pm 0.56 $ &	- &  J225430-402329 &  0.36  &  1.10  &  $-0.88 \pm 0.32 $ \\
 J224801-400542 &   - &  0.45  &  0.49  &  $-0.07 \pm 0.38 $ &	 J225434-401343 &    J225434-401343 &  7.80  &  21.09  &  $-0.78 \pm 0.01 $ \\
- &  J224801-395900 &  0.40  &  0.69  &  $-0.43 \pm 0.36 $ &	 J225435-395931 &   - &  0.71  &  0.41  &  $0.43 \pm 0.39 $ \\
- &  J224803-400513 &  $<$0.20  &  0.67  &  $<-0.95  $ &	 J225436-400531 &   - &  0.60  &  0.47  &  $0.19 \pm 0.36 $ \\
 J224806-402102 &    J224806-402101 &  0.80  &  2.52  &  $-0.90 \pm 0.16 $ &	 J225442-400353 &   - &  0.56  &  $<$0.26  &  $>0.60 $ \\
 J224809-402211 &    J224809-402212 &  1.26  &  4.23  &  $-0.95 \pm 0.10 $ &	 J225443-401147 &   - &  0.41  &  $<$0.26  &  $>0.36  $ \\
- &  J224811-402455 &  0.28  &  0.59  &  $-0.59 \pm 0.48 $ &	 J225449-400918 &    J225449-400918 &  0.86  &  1.24  &  $-0.29 \pm 0.17 $ \\
- &  J224817-400819 &  0.31  &  0.54  &  $-0.44 \pm 0.43 $ &	 J225450-401639 &   - &  0.61  &  $<$0.26  &  $>0.67  $ \\
 J224822-401808 &    J224822-401808 &  10.34  &  19.08  &  $-0.48 \pm 0.01 $ &	 J225504-400154 &    J225504-400154 &  4.59  &  9.67  &  $-0.59 \pm 0.03 $ \\
 J224827-402515 &    J224827-402515 &  0.80  &  0.58  &  $0.25 \pm 0.27 $ &	 J225505-401301 &   - &  0.37  &  0.51  &  $-0.25 \pm 0.40 $ \\
- &  J224828-395814 &  $<$0.20  &  0.67  &  $<-0.95 $ &	 J225509-402658 &    J225509-402658 &  5.18  &  8.15  &  $-0.36 \pm 0.03 $ \\
- &  J224843-400456 &  $<$0.18  &  0.72  &  $<-1.09 $ &	 J225511-401513 &    J225512-401513 &  0.58  &  0.71  &  $-0.16 \pm 0.27 $ \\
 J224850-400027 &    J224850-400027 &  1.37  &  1.10  &  $0.17 \pm 0.14 $ &	 J225515-401835 &    J225515-401835 &  0.42  &  1.07  &  $-0.73 \pm 0.29 $ \\
 J224858-402708 &    J224858-402707 &  0.50  &  0.96  &  $-0.51 \pm 0.26 $ &	 J225526-400112 &    J225526-400111 &  0.41  &  1.55  &  $-1.04 \pm 0.28 $ \\
 J224903-400946 &    J224903-400949 &  0.96  &  2.81  &  $-0.84 \pm 0.12 $ &	 J225529-401101 &    J225529-401101 &  1.09  &  1.48  &  $-0.24 \pm 0.14 $ \\
 J224906-402337 &    J224906-402337 &  2.48  &  3.01  &  $-0.15 \pm 0.06 $ & &&&&\\
&&&&&&&&\\
\hline
\end{tabular}
\end{flushleft}
\end{table*}

\begin{figure}[t]
\resizebox{\hsize}{!}{\includegraphics{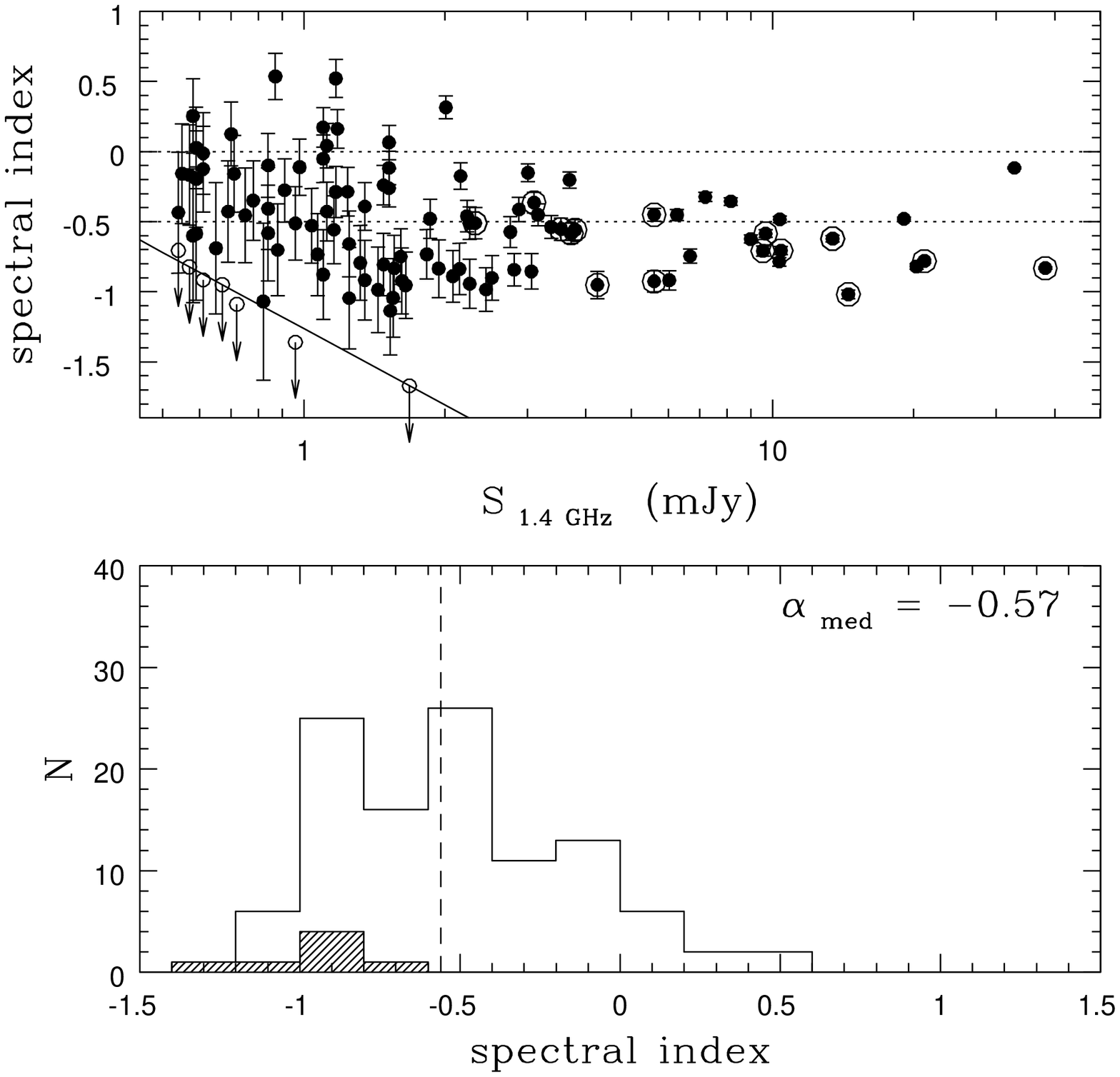}}
\caption{{\it Top:} 1.4 -- 5 GHz spectral index vs. 1.4 GHz~flux density for 
the ATESP sample. Only sources catalogued at 1.4~GHz ($S_{\rm peak}({\rm 
1.4 \, GHz})\geq 6\sigma$) are shown. Open circles with arrows 
indicate spectral index upper 
limits due to sources undetected at 5~GHz (the solid line indicates the 
$3\sigma$ detection limit). 
Horizontal dotted lines show the $\alpha = 0$ and the 
$\alpha = -0.5$ loci. Circled dots refer to sources
displaying a multiple-component/extended radio morphology (either at 1.4 or 
5~GHz, either at low or full resolution), typical of AGN--driven radio 
galaxies. {\it Bottom:} Spectral index 
distribution for the same sources shown above. Shaded histogram refer to upper 
limits. The vertical dashed line indicates the median spectral index value of
the sample ($\alpha_{\rm med} = -0.57$).} 
\label{fig:siflux20}
\end{figure}

\begin{figure}[t]
\resizebox{\hsize}{!}{\includegraphics[scale=0.6]{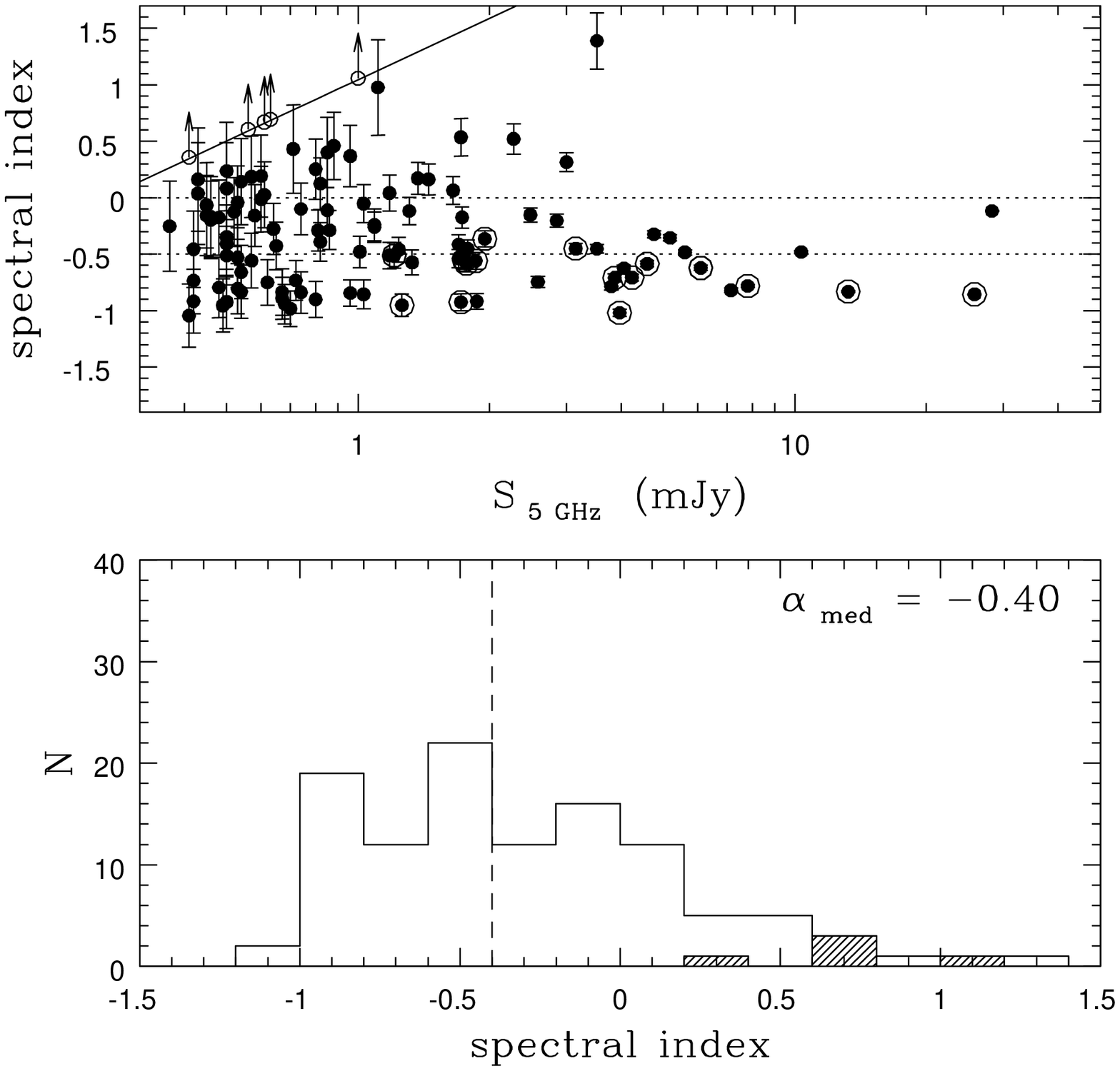}}
\caption{{\it Top:} 1.4 -- 5 GHz spectral index vs. 5~GHz flux density for the 
ATESP sample. Only sources catalogued at 5~GHz ($S_{\rm peak}({\rm 
5 \, GHz})\geq 6\sigma$) are shown. Open circles with arrows 
indicate spectral index lower 
limits due to sources undetected at 1.4 GHz (the solid line indicates 
the $3\sigma$ detection limit). 
Horizontal dotted lines indicate the $\alpha = 0$ and the 
$\alpha = -0.5$ loci. Circled dots refer to sources
displaying a multiple-component/extended radio morphology (either at 1.4 or 
5~GHz, either at low or full resolution), typical of AGN--driven radio 
galaxies. {\it Bottom:} Spectral index 
distribution for the same sources shown above. Shaded histogram refer to lower 
limits. The vertical dashed line indicates the median spectral index value of
the sample ($\alpha_{\rm med} = -0.40$).} 
\label{fig:siflux5}
\end{figure}

For the analysis of the $1.4-5$ GHz spectral index properties of the 
ATESP radio sources, we extracted from the ATESP
1.4~GHz catalogue (Prandoni et al., 2000b), the list of sources
located in the $2\times 0.5$ sq. degr. region surveyed at 5~GHz.
Of the 109 1.4~GHz sources found, 89 are catalogued also at 5~GHz, whereas 
the remaining 20 sources are catalogued only at 1.4~GHz. Similarly we have 
22 sources which are catalogued only at 5~GHz, for a total of 131 
(1.4 and/or 5~GHz) ATESP sources in the overlapping region. 
In order to exploit the whole sample of 131 sources 
for the spectral index analysis, we searched for counterparts of the 
42 ($20+22$) sources catalogued only at 1.4 (or 5 GHz) down to a 
$3\sigma$-threshold ($\sim 0.2$ mJy), by directly inspecting the low resolution
5~GHz (or the 1.4 GHz) ATESP radio mosaics, at the source position. 
This allows us to probe deeper fluxes than with the 89 ATESP sources 
catalogued at both 1.4 
and 5~GHz ($S_{\rm peak}>6\sigma \sim 0.4-0.5$ mJy). 

The result of this search is reported in 
Table~\ref{tab-si}, where, for each source, we list the 5~GHz and/or 1.4~GHz 
ATESP IAU source name (Columns 1 and 2), the 5~GHz (low resolution) and 1.4~GHz
flux density used for the spectral index derivation (Columns 3 and 4), the
$1.4 - 5$ GHz spectral index with its standard deviation (Column 5).   

We notice that flux densities reported in Table~\ref{tab-si} are 
integrated values for resolved sources and peak values for either 
point-like sources or $<6\sigma$ detections. 
For undetected sources $3\sigma_{\rm local}$ upper limits are provided.
In addition, 1.4~GHz flux densities are corrected for systematic effects (e.g. 
clean bias and radial smearing), following the recipe derived in 
Prandoni et al. 2000b (Appendix B).

In summary we have that 118 of the 131 ATESP sources are detected at both 
frequencies (down to a $3\sigma$-threshold), 5 are detected only at 5~GHz 
and 8 only at 1.4~GHz. 

Figures~\ref{fig:siflux20} and \ref{fig:siflux5} show the $1.4$--$5$ GHz
spectral index as a function of flux density, for the 109 sources catalogued 
at  1.4~GHz and for the 111 sources catalogued at 5~GHz respectively 
(see top panels). Also plotted
are the spectral index distributions at the two frequencies (bottom panels). 
As expected for higher frequency selected samples, the
median spectral index is flatter at 5~GHz ($\alpha_{\rm med}=-0.40$) 
than at 1.4~GHz ($\alpha_{\rm med}=-0.57$) and both samples show a 
flattening (much more evident at 5~GHz) going to lower flux 
densities (see Table~\ref{tab-sivalues}). 
For $S_{\mbox{1.4 GHz}}>4$ mJy, $\alpha$ shows a narrow dispersion around a 
median value of $-0.71$ ($-0.62$ at 5 GHz), as expected for standard 
synchrotron radio emission. At fainter fluxes, however, almost half ($46\%$) 
of the sources selected at 1.4 GHz and almost 2/3 ($\sim 63\%$) 
of the sources selected 
at 5 GHz  show flat spectra ($\alpha >-0.5$), with a significant fraction
($29\%$at 5 GHz) of inverted spectra ($\alpha >0$). 

It is worth noticing that, since the 5~GHz survey has been conducted several 
years later than the 1.4 GHz survey, flux variability could affect 
the spectral index derivation for a number of sources. In particular, 
it could explain the most extreme cases (like ultra-steep and/or 
ultra-inverted sources). On the other hand, the two ultra--steep sources 
($\alpha <-1.3$) present in the sample
(see Fig.~\ref{fig:siflux20}), if real, could potentially be 
associated to very high redshift galaxies. 

The flattening of the radio spectral index going from mJy to sub--mJy flux 
densities found for the ATESP--DEEP1 sample confirms on a much larger 
statistical basis (131 sources) previous results obtained 
for smaller ($50-60$ sources) samples 
(e.g. Donnelly et al 1987; Gruppioni et al. 1997) and agrees with the
finding of many flat/inverted radio spectra in $\mu$Jy samples 
(Fomalont et al. 1991; Windhorst et al 1993). On the other hand, it  
disagrees with a more recent spectral index analysis performed in the Lockman 
Hole, where the sources are claimed to have steep--spectrum down to 0.2 mJy 
(Ciliegi et al. 2003). It is worth noticing, however, that the statistics 
available to the Ciliegi et al. sample at $S_{\rm 5 \, GHz}>0.2$ mJy is very 
poor.

It is interesting to compare the spectral index statistics reported in 
Table~\ref{tab-sivalues} (upper limits are included by using the survival 
analysis methods\footnote{Package ASURV Rev. 2.1} 
presented in Feigelson \& Nelson (1985) and Isobe et al. 
(1986)) to the one obtained in samples selected 
at the same frequencies (1.4 and 5 GHz). As mentioned above, a
general agreement is found: Fomalont et al. (1991) report a
$\alpha_{med}=-0.38$ and a $f(\alpha>-0.5)=60\%$ at fluxes 
$16<S_{\rm 5~GHz}<1000$ $\mu$Jy; while Donnelly et al. (1987) report 
$\alpha_{mean}=-0.31\pm0.58$, $\alpha_{med}=-0.42$ and $f(\alpha>-0.5)=50\%$ 
at $0.4<S_{\rm 5~GHz}<1.2$ mJy. On the other hand, a somewhat 
steeper behaviour 
was found also by Donnelly et al. (1987) at 1.4 GHz: 
$\alpha_{mean}=-0.80\pm0.49$, $\alpha_{med}=-0.76$ and $f(\alpha>-0.5)=22\%$ 
at $0.5<S_{\rm 1.4~GHz}<3$ mJy, which was interpreted as due to a significantly
different composition of the faint radio population depending on the selection
frequency. 

Indeed the ATESP 1.4 and 5~GHz selected samples have in common only 2/3 of 
the sources (89/131). This can explain not only the somewhat different 
spectral properties of the two samples at low flux densities ($S<4$ mJy), 
but also why a significant flattening of source counts in 1.4~GHz-selected 
samples at $S\la 1$ mJy does not necessarily result in a 
flattening of the 5~GHz source counts in the flux range covered by the 
ATESP sample. By assuming as reference $\alpha = -0.53$ (the median spectral 
index found for the sub-mJy 1.4~GHz selected ATESP sources, see 
Tab.~\ref{tab-sivalues}), we expect the flattening to occur at 
$S_{\rm 5 \, GHz}\sim 0.5$ mJy, that is very close to the flux density limit 
of the 5~GHz ATESP survey. 

As a final remark we notice that radio sources with a 
multiple--component/extended radio morphology, typical of classical 
AGN--driven radio galaxies, are both shown as circled dots in 
Figs.~\ref{fig:siflux20} and \ref{fig:siflux5}. 
It is evident that such sources are characterized by steep synchrotron
radio spectra and are mainly found at mJy flux densities. On the other hand
unresolved multiple--component sources can be present also at lower fluxes, 
where most of the sources could not be successfully deconvolved, due to the 
poorer deconvolution efficiency (see \S~\ref{sec-sizes}). Higher
spatial resolution radio data are needed to assess whether radio emission
in sub-mJy flat sources is triggered by active nuclei or large-scale 
star formation.  

\begin{table}
\caption{Spectral index statistics}
\label{tab-sivalues}
\begin{tabular}{l|ccccc}
\hline
\hline
Flux range & N & $\alpha_{\rm mean}$ & $\alpha_{\rm med}$ & $N_{\alpha>-0.5}$ & $N_{\alpha>0}$ \\
\hline
1.4~GHz &&&&&\\
\hline
Any flux & 109 &  \scriptsize{$-0.56\pm0.04$}  & $-0.57$ & 47 (43\%) & 10 (9\%) \\
$S>4$ mJy & 22 &  \scriptsize{$-0.66\pm0.05$}  & $-0.71$ & 7 (32\%)  & - \\
$S\leq 4$ mJy & 87 & \scriptsize{$-0.53\pm0.05$}  & $-0.53$ & 40 (46\%)  & 10 (11\%)\\
\hline
5~GHz &&&&&\\
\hline
Any flux & 111 &  \scriptsize{$-0.28\pm0.05$}  & $-0.40$ & 67 (60\%)  & 28 (25\%) \\
$S>4$ mJy & 13 &  \scriptsize{$-0.58\pm0.06$}  & $-0.62$ & 5 (38\%)  & - \\
$S\leq 4$ mJy & 98 & \scriptsize{$-0.24\pm0.06$}  & $-0.29$ & 62 (63\%)  & 28 (29\%)\\
\hline
\end{tabular}
\end{table}

\section{Summary}\label{sec-summary}

We used the ATCA to follow-up at 5 GHz part of the region previously covered
by the sub--mJy ATESP 1.4 GHz survey (Prandoni et al. 2000a,b). 
In particular, we imaged a 1~sq.~degr. area where, in addition to the 1.4 GHz
information, extensive $UBVRI$ deep optical imaging is 
available (Mignano et al. 2006). 
The 5 GHz survey was designed to have uniform sensitivity over the 
entire region observed.

In order to be able to measure the 1.4 - 5 GHz spectral index,
we produced 5 GHz images at exactly the same resolution (same size in pixels 
and restoring beam) as the 1.4 GHz mosaics (see Prandoni 
et al. \cite{Prandoni00a}). On the other hand, full 
resolution ($\sim 2\arcsec$) mosaics were also produced. As expected, the 
noise level obtained, $\sim 70$ $\mu$Jy, is fairly uniform within each 
mosaic and from mosaic to mosaic. 

We used the low resolution images for the source extraction: we produced a 
catalogue of 111 radio sources, complete down to $S\sim 0.4$ mJy ($6\sigma$).
The full resolution information was used to assess {\bf size} and 
radio morphology 
of the catalogued sources. A detailed analysis of the source sizes 
shows that the ATESP 5 GHz sources can be satisfactorily described by the
angular size distribution proposed by Windhorst et al. (1990), at least down
to $S>0.7-1$ mJy. Below such limit, the high number of sources which remain 
unresolved even at full resolution prevents us from a reliable analysis. 
We can only argue that our upper limits {\bf ($<1-2\arcsec$)} 
are consistent with 
the result of Fomalont et al. (1991), who find $\Theta < 1.5\arcsec$ for 
sources with $0.06< S(\rm{5GHz}) < 0.8$ mJy. Also 
consistent is the result of Bondi et al. (2003) who find 
{\bf $\Theta_{med} = 1.\arcsec 8 \pm 0.\arcsec 2$.} 
We notice that Bondi et al. suggest a steeper 
distribution {\bf than} the one of Windhorst et al. for sources 
$0.4<S(\rm{1.4 GHz})<1$ mJy.   

We have derived the $\log{N} - \log{S}$ relation for the ATESP 5 GHz 
catalogue. The possible causes of incompleteness at the faint end of the 
source counts have been taken into account, with particular respect to
resolution effects (resolution bias). 
The ATESP 5 GHz counts are consistent with the counts derived from 
other 5 GHz surveys and improve significantly the statistics of the counts 
in the flux range $0.4 - 1$ mJy. The ATESP 5 GHz counts 
do not show evidence of flattening down to the survey limit. 

The $1.4-5$~GHz spectral index was derived for all the ATESP 
radio sources present in the $2\times 0.5$ sq. degr. region surveyed at both
1.4 and 5~GHz. A flattening of the spectral index 
with decreasing flux densities was found, which is particularly significant
for the 5~GHz selected sample.
At mJy level we have mostly steep-spectrum ($\alpha\sim -0.7$) 
synchrotron radio sources, 
while at sub-mJy flux densities we have a composite 
population, with $\sim 60\%$ of the 5~GHz sources showing flat spectra 
and a significant fraction ($\sim 30\%$ at 5~GHz) of inverted-spectrum 
sources.   
The spectral index flattening of the 1.4~GHz selected sample is somewhat 
milder, with $\alpha_{\rm med}=-0.53$ at sub-mJy flux densities against 
a value of $\alpha_{\rm med}=-0.29$ found for the 5~GHz sources. 

We notice that, when we take into account the spectral index properties
of the ATESP 1.4~GHz selected sample, as inferred by our 1.4/5~GHz 
measurements, we can expect the 5~GHz source counts to flatten at flux 
densities $\sim 0.5$ mJy, that is very close to the flux density limit of
the 5~GHz ATESP survey. This explains why we do not yet see evidence
of such a flattening in the ATESP 5~GHz source counts.  

Particularly interesting is the 
possibility of combining the spectral index information with other 
observational properties to infer the nature of the faint 
radio population, with particular respect to flat/inverted--spectrum sources, 
and assess the physical processes triggering the radio
emission in those sources. This kind of analysis needs information about
redshifts and types of the galaxies hosting the radio sources. 
A detailed radio/optical study of our sample is possible, thanks to the 
extensive optical coverage available to it (see Sect.~\ref{sec-20cm}), and 
is the subject of the second paper of this series 
(Mignano et al., in prep.).

\begin{acknowledgements}
The Australia Telescope is funded by the Commonwealth of Australia for 
operation as a National Facility managed by CSIRO.
\end{acknowledgements}

\clearpage

\begin{sidewaystable*}
\begin{flushleft}
\caption[]{The 5 GHz Catalogue}
\label{tab-5ghzlhrcat}
\scriptsize
\begin{tabular}{l|ccrrrrrl|c|ccrrrrrl|c}
\hline\hline
\multicolumn{1}{c|}{} &
\multicolumn{9}{c|}{low resolution parameters} 
& \multicolumn{9}{c}{full resolution parameters}\\
\hline
 \multicolumn{1}{c|}{IAU Name}   & \multicolumn{1}{c}{R.A.}
& \multicolumn{1}{c}{DEC.} & \multicolumn{1}{c}{$S_{\rm peak}$} & 
\multicolumn{1}{c}{$S_{\rm total}$} & \multicolumn{1}{c}{$\Theta_{\rm maj}$} & 
\multicolumn{1}{c}{$\Theta_{\rm min}$} & \multicolumn{1}{c}{P.A.} &
\multicolumn{1}{c|}{} & \multicolumn{1}{c|}{$\sigma_{\rm local}$} 
&  \multicolumn{1}{c}{R.A.}
& \multicolumn{1}{c}{DEC.} & \multicolumn{1}{c}{$S_{\rm peak}$} & 
\multicolumn{1}{c}{$S_{\rm total}$} & \multicolumn{1}{c}{$\Theta_{\rm maj}$} & 
\multicolumn{1}{c}{$\Theta_{\rm min}$} & \multicolumn{1}{c}{P.A.} &
\multicolumn{1}{c|}{} & \multicolumn{1}{c}{$\sigma_{\rm local}$}     \\
 \multicolumn{1}{c|}{} & \multicolumn{2}{c}{(J2000)} 
& \multicolumn{2}{c}{mJy} 
& \multicolumn{2}{c}{arcsec} & \multicolumn{1}{c}{degr.} 
& \multicolumn{1}{c|}{} & \multicolumn{1}{c|}{$\mu$Jy} 
& \multicolumn{2}{c}{(J2000)}  
& \multicolumn{2}{c}{mJy} & \multicolumn{2}{c}{arcsec} 
& \multicolumn{1}{c}{degr.} & \multicolumn{1}{c|}{} 
& \multicolumn{1}{c}{$\mu$Jy}\\
\hline
&&&&&&&&&&&&&&&&&& \\
  ATESP5 J224502-400415 & 22 45 02.86 & -40 04 15.4 & 0.65 & 0.62 & 0.00 & 0.00 & 0.0 & S  & 67 & 22 45 02.83 & -40 04 14.7 & 0.43 & 0.60 & 1.86 & 1.14 & -42.9 & S* & 69 \\   
  ATESP5 J224509-400622 & 22 45 09.40 & -40 06 22.8 & 0.70 & 0.67 & 0.00 & 0.00 & 0.0 & S  & 61 & 22 45 09.38 & -40 06 23.0 & 0.53 & 0.72 & 2.50 & 0.35 & -27.9 & S & 67 \\   
  ATESP5 J224510-401655 & 22 45 10.00 & -40 16 55.2 & 0.46 & 0.66 & 0.00 & 0.00 & 0.0 & S* & 75 & 22 45 09.99 & -40 16 56.8 & 0.34 & &&&& D & 74 \\ 
  ATESP5 J224513-400052 & 22 45 13.46 & -40 00 52.3 & 0.46 & 0.44 & 0.00 & 0.00 & 0.0 & S & 72 & 22 45 13.51 & -40 00 51.1 & 0.33 & &&& & D& 76  \\ 
&&&&&&&&&&&&&&&&&&  \\
  ATESP5 J224516-401807 & 22 45 16.70 & -40 18 07.8 & 1.62 & 3.87 & 18.38 & 0.00 & 0.0 & M &&&&&&&&&& \\
  ATESP5 J224516-401807A & 22 45 15.86 & -40 17 58.5 & 0.69 & 1.04 & 9.03 & 0.00 & -81.2 & S & 76 & 22 45 16.01 & -40 17 58.5 & 0.37 &  &  &  &  & D & 72  \\ 
  ATESP5 J224516-401807B & 22 45 17.01 & -40 18 11.3 & 1.62 & 2.83 & 12.58 & 4.81 & -26.4 & S &76 & 22 45 16.70 & -40 18 05.9 & 0.54 & 0.59 & 0.00 & 0.00 & 0.0 & S($^a$) & 72   \\ 
&&&&&&&&&& 22 45 17.13 & -40 18 13.7 & 0.54 & 2.08 & 11.40 & 5.27 & -11.7 & E &
72   \\ 
&&&&&&&&&&&&&&&&&&   \\
 ATESP5 J224518-401001 & 22 45 18.70 & -40 10 01.6 & 5.60 & 5.65 & 0.00 & 0.00 & 0.0 & S & 59 & 22 45 18.69 & -40 10 01.7 & 4.54 & 5.03 & 1.68 & 0.00 & 32.5 & S & 85  \\ 
 ATESP5 J224530-401141 & 22 45 30.30 & -40 11 41.8 & 0.88 & 0.75 & 0.00 & 0.00 & 0.0 & S & 69 & 22 45 30.30 & -40 11 41.3 & 0.67 & 0.73 & 0.00 & 0.00 & 0.0 & S &67  \\ 
 ATESP5 J224533-402014 & 22 45 33.03 & -40 20 14.6 & 0.53 & 0.53 & 0.00 & 0.00 & 0.0 & S & 74 & 22 45 33.10 & -40 20 15.7 & 0.39 & &&& & D & 66  \\  
 ATESP5 J224534-401337 & 22 45 34.37 & -40 13 37.2 & 1.17 & 1.86 & 10.52 & 4.12 & 32.4 & S & 71 & 22 45 34.49 & -40 13 34.1 & 0.38 & 1.12 & 8.32 & 5.19 & 40.8 & DE($^b$) & 69  \\
 ATESP5 J224534-400049 & 22 45 34.33 & -40 00 49.4 & 1.54 & 1.87 & 7.66 & 1.31 & -2.4 & S & 73 & 22 45 34.31 & -40 00 47.6 & 0.98 & 1.26 & 1.83 & 1.10 & -9.9 & S & 94 \\  
 ATESP5 J224547-400324 & 22 45 47.88 & -40 03 24.2 & 26.86 & 28.28 & 3.79 & 0.00 & 46.1 & S & 78 & 22 45 47.86 & -40 03 24.2 & 18.89 & 21.44 & 1.55 & 0.00 & 34.8 & S & 184  \\ 
 ATESP5 J224550-402021 & 22 45 50.47 & -40 20 21.8 & 0.60 & 1.75 & 28.89 & 20.23 & -71.2 & E & 74 & 22 45 50.55 & -40 20 26.3 & 0.25 & &&&& D($^c$) & 73  \\
 ATESP5 J224551-401618 & 22 45 51.30 & -40 16 18.8 & 0.57 & 0.54 & 0.00 & 0.00 & 0.0 & S & 73 & 22 45 51.29 & -40 16 20.1 & 0.40 & &&& & D & 76 \\ 
 ATESP5 J224557-400934 & 22 45 57.79 & -40 09 34.1 & 0.42 & 0.52 & 0.00 & 0.00 & 0.0 & S & 66 & 22 45 57.83 & -40 09 34.2 & 0.29 & &&& & D & 80 \\ 
&&&&&&&&&&&&&&&&&&  \\
 ATESP5 J224559-402122 & 22 45 59.03 & -40 21 22.0 & 1.09 & 1.72 & 20.00 & 0.00 & 0.0 & M &&&&&&&&&& \\
 ATESP5 J224559-402122A & 22 45 57.99 & -40 21 22.8 & 0.70 & 0.70 & 0.00 & 0.00 & 0.0 & S & 73 & 22 45 57.97 & -40 21 23.7 & 0.61 & 0.69 & 0.00 & 0.00 & 0.0 & S & 73 \\ 
 ATESP5 J224559-402122B & 22 45 59.74 & -40 21 21.5 & 1.09 & 1.02 & 0.00 & 0.00 & 0.0 & S & 73 & 22 45 59.73 & -40 21 21.9 & 0.81 & 0.95 & 2.19 & 0.00 & -11.6 & S & 73 \\ 
&&&&&&&&&&&&&&&&&&  \\
 ATESP5 J224601-401502 & 22 46 01.50 & -40 15 02.9 & 2.58 & 2.55 & 0.00 & 0.00 & 0.0 & S & 69 & 22 46 01.49 & -40 15 02.4 & 2.11 & 2.44 & 1.26 & 0.58 & 64.1 & S & 73 \\ 
 ATESP5 J224608-400414 & 22 46 08.44 & -40 04 14.5 & 0.61 & 0.74 & 0.00 & 0.00 & 0.0 & S & 68 & 22 46 08.47 & -40 04 15.4 & 0.48 & &&&& D & 83  \\  
 ATESP5 J224613-401132 & 22 46 13.17 & -40 11 32.2 & 1.24 & 1.28 & 0.00 & 0.00 & 0.0 & S & 70 & 22 46 13.15 & -40 11 32.4 & 0.95 & 0.97 & 0.00 & 0.00 & 0.0 & S & 63  \\ 
 ATESP5 J224623-400854 & 22 46 23.35 & -40 08 54.7 & 0.68 & 0.69 & 0.00 & 0.00 & 0.0 & S & 67 & 22 46 23.33 & -40 08 56.3 & 0.60 & 0.70 & 0.00 & 0.00 & 0.0 & S* & 80  \\ 
&&&&&&&&&&&&&&&&&&   \\
\hline
\end{tabular}
\end{flushleft}
\end{sidewaystable*}

\addtocounter{table}{-1}
\clearpage

\begin{sidewaystable*}
\begin{flushleft}
\caption[]{The 5 GHz Catalogue (continued)}
\scriptsize
\begin{tabular}{l|ccrrrrrl|c|ccrrrrrl|c}
\hline\hline
\multicolumn{1}{c|}{} &
\multicolumn{9}{c|}{low resolution parameters} 
& \multicolumn{9}{c}{full resolution parameters}\\
\hline
 \multicolumn{1}{c|}{IAU Name}   & \multicolumn{1}{c}{R.A.}
& \multicolumn{1}{c}{DEC.} & \multicolumn{1}{c}{$S_{\rm peak}$} & 
\multicolumn{1}{c}{$S_{\rm total}$} & \multicolumn{1}{c}{$\Theta_{\rm maj}$} & 
\multicolumn{1}{c}{$\Theta_{\rm min}$} & \multicolumn{1}{c}{P.A.} &
\multicolumn{1}{c|}{} & \multicolumn{1}{c|}{$\sigma_{\rm local}$} 
&  \multicolumn{1}{c}{R.A.}
& \multicolumn{1}{c}{DEC.} & \multicolumn{1}{c}{$S_{\rm peak}$} & 
\multicolumn{1}{c}{$S_{\rm total}$} & \multicolumn{1}{c}{$\Theta_{\rm maj}$} & 
\multicolumn{1}{c}{$\Theta_{\rm min}$} & \multicolumn{1}{c}{P.A.} &
\multicolumn{1}{c|}{} & \multicolumn{1}{c}{$\sigma_{\rm local}$} \\
 \multicolumn{1}{c|}{} & \multicolumn{2}{c}{(J2000)} & \multicolumn{2}{c}{mJy} 
& \multicolumn{2}{c}{arcsec} & \multicolumn{1}{c}{degr.} 
& \multicolumn{1}{c|}{} & \multicolumn{1}{c|}{$\mu$Jy} 
& \multicolumn{2}{c}{(J2000)}  
& \multicolumn{2}{c}{mJy} & \multicolumn{2}{c}{arcsec} 
& \multicolumn{1}{c}{degr.} & \multicolumn{1}{c|}{} 
& \multicolumn{1}{c}{$\mu$Jy}\\
\hline
&&&&&&&&&&&&&&&&&& \\
 ATESP5 J224628-401207 & 22 46 28.46 & -40 12 07.4 & 0.82 & 0.94 & 0.00 & 0.00 & 0.0 & S & 68 & 22 46 28.46 & -40 12 07.3 & 0.57 & 0.71 & 0.00 & 0.00 & 0.0 & S & 76  \\ 
 ATESP5 J224632-400319 & 22 46 32.88 & -40 03 19.0 & 0.72 & 0.75 & 0.00 & 0.00 & 0.0 & S & 69 & 22 46 32.89 & -40 03 19.1 & 0.42 & 0.88 & 3.41 & 1.86 & -88.3 & S & 69 \\  
&&&&&&&&&&&&&&&&&&  \\
 ATESP5 J224646-400829 & 22 46 46.52 & -40 08 29.6 & 8.93 & 13.25 & 11.65 & 0.00 & 32.5 & S & 70 & 22 46 46.26 & -40 08 33.8 & 2.79 & 4.62 & 3.19 & 1.57 & 23.4 & S($^d$) & 96 \\ 
&&&&&&&&&& 22 46 46.64 & -40 08 28.2 & 3.70 & 7.26 & 5.36 & 1.09 & 7.4 & S 
& 96 \\ 
&&&&&&&&&&&&&&&&&&   \\
 ATESP5 J224647-401220 & 22 46 47.66 & -40 12 20.7 & 0.81 & 0.66 & 0.00 & 0.00 & 0.0 & S & 69 & 22 46 47.65 & -40 12 21.5 & 0.64 & 0.83 & 2.84 & 0.00 & 16.0 & S & 73 \\  
 ATESP5 J224654-400107 & 22 46 54.53 & -40 01 07.2 & 1.17 & 3.15 & 36.66 & 27.04 & 89.8 & E & 73 & 22 46 54.55 & -40 01 06.1 & 0.49 & 0.70 & 2.86 & 0.67 & 36.9 & S*($^e$)  & 73  \\
 ATESP5 J224701-402646 & 22 47 01.48 & -40 26 46.1 & 1.08 & 1.33 & 5.72 & 1.88 & 48.5 & S  & 79 & 22 47 01.48 & -40 26 45.9 & 0.74 & 1.00 & 2.43 & 1.03 & -15.8 & S & 74   \\  
 ATESP5 J224702-400948 & 22 47 02.08 & -40 09 48.1 & 0.60 & 0.57 & 0.00 & 0.00 & 0.0 & S & 70 & 22 47 02.03 & -40 09 48.6 & 0.41 & &&& & D & 75  \\  
 ATESP5 J224707-400616 & 22 47 07.50 & -40 06 16.6 & 1.70 & 1.76 & 0.00 & 0.00 & 0.0 & S & 71 & 22 47 07.51 & -40 06 16.6 & 1.33 & 1.51 & 1.88 & 0.00 & 40.3 & S & 73   \\  
 ATESP5 J224714-401453 & 22 47 14.21 & -40 14 53.9 & 2.89 & 3.52 & 6.43 & 0.00 & 53.2 & S & 71 & 22 47 14.19 & -40 14 54.4 & 1.45 & 3.27 & 4.54 & 1.74 & 35.4 & S& 76  \\ 
 ATESP5 J224714-402400 & 22 47 14.71 & -40 24 00.2 & 0.57 & 0.47 & 0.00 & 0.00 & 0.0 & S & 71 & 22 47 14.70 & -40 24 01.1 & 0.52 & 0.44 & 0.00 & 0.00 & 0.0 & S & 78 \\  
 ATESP5 J224719-400141 & 22 47 19.56 & -40 01 41.0 & 0.48 & 0.56 & 0.00 & 0.00 & 0.0 & S* & 71 & 22 47 19.53 & -40 01 41.9 & 0.53 & 0.49 & 0.00 & 0.00 & 0.0 & S & 69 \\ 
 ATESP5 J224719-401530 & 22 47 19.61 & -40 15 30.3 & 0.74 & 0.56 & 0.00 & 0.00 & 0.0 & S & 73 & 22 47 19.61 & -40 15 30.7 & 0.59 & 0.61 & 0.00 & 0.00 & 0.0 & S & 73 \\   
 ATESP5 J224721-402043 & 22 47 21.66 & -40 20 43.9 & 0.50 & 0.57 & 0.00 & 0.00 & 0.0 & S & 74 & 22 47 21.60 & -40 20 42.5 & 0.26 & &&&& D & 76 \\ 
 ATESP5 J224724-395909 & 22 47 24.75 & -39 59 09.2 & 1.18 & 1.45 & 6.49 & 0.22 & 38.5 & S & 71 & 22 47 24.77 & -39 59 08.9 & 1.00 & 1.02 & 0.00 & 0.00 & 0.0 & S & 73  \\ 
 ATESP5 J224727-402751 & 22 47 27.25 & -40 27 51.0 & 0.67 & 0.59 & 0.00 & 0.00 & 0.0 & S & 77 & 22 47 27.22 & -40 27 50.4 & 0.61 & 0.52 & 0.00 & 0.00 & 0.0 & S & 78 \\  
 ATESP5 J224727-401926 & 22 47 27.31 & -40 19 26.6 & 1.11 & 1.12 & 0.00 & 0.00 & 0.0 & S & 73 & 22 47 27.30 & -40 19 26.3 & 0.93 & 1.02 & 0.00 & 0.00 & 0.0 & S &  72 \\  
 ATESP5 J224729-402000 & 22 47 29.42 & -40 20 00.5 & 0.50 & 0.41 & 0.00 & 0.00 & 0.0 & S & 73 & 22 47 29.52 & -40 20 01.3 & 0.35 & &&& & D & 75  \\  
 ATESP5 J224731-400527 & 22 47 31.58 & -40 05 27.8 & 0.43 & 0.34 & 0.00 & 0.00 & 0.0 & S & 67  & 22 47 31.63 & -40 05 28.7 & 0.28 & &&&& D & 76  \\  
 ATESP5 J224732-401442 & 22 47 32.94 & -40 14 42.7 & 6.07 & 7.14 & 7.14 & 2.51 & 9.2 & S & 75  & 22 47 32.92 & -40 14 42.3 & 3.48 & 6.05 & 3.70 & 1.46 & -22.4 & S & 79\\   
  ATESP5 J224735-402321 & 22 47 35.70 & -40 23 21.8 & 1.18 & 1.16 & 0.00 & 0.00 & 0.0 & S & 75 & 22 47 35.72 & -40 23 21.7 & 1.08 & 1.27 & 2.37 & 0.00 & 17.4 & S & 74 \\ 
  ATESP5 J224740-401821 & 22 47 40.49 & -40 18 21.7 & 0.74 & 0.62 & 0.00 & 0.00 & 0.0 & S & 76 & 22 47 40.53 & -40 18 20.5 & 0.50 & 0.93 & 5.65 & 0.14 & 8.2 & S* & 77 \\ 
  ATESP5 J224741-400442 & 22 47 41.13 & -40 04 42.6 & 0.52 & 0.48 & 0.00 & 0.00 & 0.0 & S  & 67 & 22 47 41.15 & -40 04 43.4 & 0.39 & &&& & D & 71 \\ 
&&&&&&&&&&&&&&&&&&  \\
  ATESP5 J224748-401324 & 22 47 48.74 & -40 13 24.5 & 2.97 & 3.97 & 6.46 & 1.95 & 76.9 & S & 72 & 22 47 48.54 & -40 13 24.3 & 1.12 & 1.81 & 2.65 & 1.59 & -44.2 & S($^d$) & 74 \\ 
&&&&&&&&&& 22 47 48.92 & -40 13 25.0 & 1.07 & 1.58 & 2.40 & 1.53 & 15.2 & S* & 74  \\ 
&&&&&&&&&&&&&&&&& \\
\hline
\end{tabular}
\end{flushleft}
\end{sidewaystable*}

\addtocounter{table}{-1}
\clearpage

\begin{sidewaystable*}
\begin{flushleft}
\caption[]{The 5 GHz Catalogue (continued)}
\scriptsize
\begin{tabular}{l|ccrrrrrl|c|ccrrrrrl|c}
\hline\hline
\multicolumn{1}{c|}{} &
\multicolumn{9}{c|}{low resolution parameters} 
& \multicolumn{9}{c}{full resolution parameters}\\
\hline
 \multicolumn{1}{c|}{IAU Name}   & \multicolumn{1}{c}{R.A.}
& \multicolumn{1}{c}{DEC.} & \multicolumn{1}{c}{$S_{\rm peak}$} & 
\multicolumn{1}{c}{$S_{\rm total}$} & \multicolumn{1}{c}{$\Theta_{\rm maj}$} & 
\multicolumn{1}{c}{$\Theta_{\rm min}$} & \multicolumn{1}{c}{P.A.} &
\multicolumn{1}{c|}{} & \multicolumn{1}{c|}{$\sigma_{\rm local}$} 
&  \multicolumn{1}{c}{R.A.}
& \multicolumn{1}{c}{DEC.} & \multicolumn{1}{c}{$S_{\rm peak}$} & 
\multicolumn{1}{c}{$S_{\rm total}$} & \multicolumn{1}{c}{$\Theta_{\rm maj}$} & 
\multicolumn{1}{c}{$\Theta_{\rm min}$} & \multicolumn{1}{c}{P.A.} &
\multicolumn{1}{c|}{} & \multicolumn{1}{c}{$\sigma_{\rm local}$} \\
 \multicolumn{1}{c|}{} & \multicolumn{2}{c}{(J2000)} 
& \multicolumn{2}{c}{mJy} 
& \multicolumn{2}{c}{arcsec} & \multicolumn{1}{c}{degr.} 
& \multicolumn{1}{c|}{} & \multicolumn{1}{c|}{$\mu$Jy} 
& \multicolumn{2}{c}{(J2000)}  
& \multicolumn{2}{c}{mJy} & \multicolumn{2}{c}{arcsec} 
& \multicolumn{1}{c}{degr.} & \multicolumn{1}{c|}{} 
& \multicolumn{1}{c}{$\mu$Jy}\\
\hline
&&&&&&&&&&&&&&&&&&  \\
  ATESP5 J224750-400148 & 22 47 50.03 & -40 01 48.6 & 2.94 & 6.09 & 56.13 & 21.93 & 53.2 & E & 72 & 22 47 50.11 & -40 01 48.6 & 1.15 & 3.18 & 5.00 & 2.77 & 26.5 & S($^f$) & 71    \\
  ATESP5 J224753-400455 & 22 47 53.73 & -40 04 55.1 & 0.67 & 0.68 & 0.00 & 0.00 & 0.0 & S & 70 & 22 47 53.65 & -40 04 56.3 & 0.44 & 0.78 & 3.51 & 0.00 & 69.4 & S* & 68 \\ 
  ATESP5 J224801-400542 & 22 48 01.06 & -40 05 42.7 & 0.45 & 0.57 & 0.00 & 0.00 & 0.0 & S & 72 & 22 48 01.12 & -40 05 42.9 & 0.30 & &&&& D & 70  \\  
  ATESP5 J224806-402102 & 22 48 06.64 & -40 21 02.3 & 0.80 & 0.69 & 0.00 & 0.00 & 0.0 & S & 72 & 22 48 06.62 & -40 21 01.6 & 0.72 & 0.61 & 0.00 & 0.00 & 0.0 & S & 78 \\ 
  ATESP5 J224809-402211 & 22 48 09.93 & -40 22 11.9 & 1.26 & 1.39 & 0.00 & 0.00 & 0.0 & S & 69 & 22 48 09.87 & -40 22 11.5 & 0.49 & 2.01 & 9.91 & 5.72 & 8.7 & E & 75  \\ 
  ATESP5 J224822-401808 & 22 48 22.12 & -40 18 08.2 & 10.26 & 10.34 & 3.02 & 0.00 & 45.5 & S & 72 & 22 48 22.12 & -40 18 08.3 & 8.70 & 9.20 & 1.05 & 0.29 & 8.1 & S & 86  \\ 
  ATESP5 J224827-402515 & 22 48 27.20 & -40 25 15.7 & 0.80 & 0.72 & 0.00 & 0.00 & 0.0 & S & 66 & 22 48 27.24 & -40 25 15.7 & 0.70 & 0.75 & 0.00 & 0.00 & 0.0 & S & 68   \\ 
  ATESP5 J224850-400027 & 22 48 50.54 & -40 00 27.4 & 1.37 & 1.41 & 0.00 & 0.00 & 0.0 & S & 63 & 22 48 50.53 & -40 00 27.5 & 1.32 & 1.36 & 0.79 & 0.00 & 6.0 & S & 66   \\ 
  ATESP5 J224858-402708 & 22 48 58.56 & -40 27 08.2 & 0.50 & 0.52 & 0.00 & 0.00 & 0.0 & S & 66 & 22 48 58.52 & -40 27 08.4 & 0.30 &  &  &  & & D & 66 \\ 
  ATESP5 J224903-400946 & 22 49 03.36 & -40 09 46.6 & 0.67 & 0.96 & 11.87 & 0.00 & 39.5 & S & 60 & 22 49 03.46 & -40 09 45.9 & 0.53 & 0.57 & 0.00 & 0.00 & 0.0 & S & 66   \\ 
  ATESP5 J224906-402337 & 22 49 06.73 & -40 23 37.5 & 2.48 & 2.36 & 0.00 & 0.00 & 0.0 & S & 66 & 22 49 06.69 & -40 23 37.7 & 2.21 & 2.33 & 1.01 & 0.28 & 10.6 & S & 72  \\   
  ATESP5 J224919-400037 & 22 49 19.35 & -40 00 37.2 & 0.64 & 0.54 & 0.00 & 0.00 & 0.0 & S & 66  & 22 49 19.34 & -40 00 37.9 & 0.73 & 0.80 & 0.00 & 0.00 & 0.0 & S & 72 \\ 
  ATESP5 J224932-395801 & 22 49 32.07 & -39 58 01.8 & 0.45 & 0.33 & 0.00 & 0.00 & 0.0 & S & 71 & 22 49 32.09 & -39 58 01.3 & 0.41 &  &  &  &  & D & 75 \\  
  ATESP5 J224935-400816 & 22 49 35.21 & -40 08 16.9 & 0.61 & 0.82 & 7.55 & 4.04 & 17.9 & S & 63 & 22 49 35.23 & -40 08 16.8 & 0.57 & 0.55 & 0.00 & 0.00 & 0.0 & S & 66  \\ 
  ATESP5 J224948-395918 & 22 49 48.08 & -39 59 18.9 & 1.72 & 1.67 & 0.00 & 0.00 & 0.0 & S & 71 & 22 49 48.08 & -39 59 19.7 & 1.86 & 1.94 & 1.04 & 0.00 & 4.5 & S & 75  \\  
  ATESP5 J224951-402035 & 22 49 51.23 & -40 20 35.4 & 0.50 & 0.60 & 0.00 & 0.00 & 0.0 & S & 66 & 22 49 51.18 & -40 20 32.6 & 0.29 &  &  &  &  & D & 71  \\   
  ATESP5 J224958-395855 & 22 49 58.26 & -39 58 55.4 & 1.65 & 1.77 & 0.00 & 0.00 & 0.0 & S & 72 & 22 49 58.27 & -39 58 55.5 & 1.46 & 1.62 & 1.33 & 0.26 & 23.8 & S & 90  \\   
  ATESP5 J225004-402412 & 22 50 04.43 & -40 24 12.4 & 1.78 & 1.76 & 0.00 & 0.00 & 0.0 & S & 66 & 22 50 04.42 & -40 24 12.5 & 1.50 & 1.63 & 1.27 & 0.45 & 0.2 & S & 76  \\   
  ATESP5 J225008-400425 & 22 50 08.82 & -40 04 25.5 & 1.49 & 1.70 & 4.86 & 2.12 & -43.6 & S & 70 & 22 50 08.81 & -40 04 25.3 & 1.33 & 1.47 & 0.94 & 0.41 & 64.2 & S & 69   \\  
  ATESP5 J225028-400333 & 22 50 28.92 & -40 03 33.3 & 0.42 & 0.40 & 0.00 & 0.00 & 0.0 & S & 68 & 22 50 28.97 & -40 03 32.0 & 0.35 & &&& & D & 73  \\    
&&&&&&&&&&&&&&&&&&  \\
  ATESP5 J225034-401936 & 22 50 34.61 & -40 19 36.3 & 15.03 & 25.78 & 13.10 & 0.00 & 0.0 & M  &&&&&&&&&&  \\  
  ATESP5 J225034-401933A & 22 50 34.28 & -40 19 33.8 & 15.03 & 16.84 & 4.11 & 2.59 & -39.1 & S & 68 & 22 50 34.26 & -40 19 33.5 & 8.63 & 13.34 & 2.48 & 1.71 & 16.6 & S & 116   \\  
  ATESP5 J225035-401941B & 22 50 35.24 & -40 19 41.0 & 8.02 & 8.94 & 3.82 & 2.85 & -29.1 & S & 68 & 22 50 35.28 & -40 19 41.2 & 4.40 & 6.55 & 2.22 & 1.69 & 11.5 & S & 116  \\     
&&&&&&&&&&&&&&&&&& \\
  ATESP5 J225048-400147 & 22 50 48.04 & -40 01 47.0 & 0.96 & 0.75 & 0.00 & 0.00 & 0.0 & S & 66 & 22 50 48.05 & -40 01 47.0 & 0.93 & 1.03 & 1.14 & 0.00 & -85.6 & S & 66  \\  
  ATESP5 J225056-402254 & 22 50 56.67 & -40 22 54.6 & 0.43 & 0.40 & 0.00 & 0.00 & 0.0 & S & 69 & 22 50 56.65 & -40 22 55.0 & 0.32 &  &  &  &  & D & 70 \\ 
  ATESP5 J225056-400033 & 22 50 56.72 & -40 00 33.2 & 2.12 & 2.27 & 3.87 & 1.72 & 4.0 & S & 65 & 22 50 56.74 & -40 00 33.2 & 1.90 & 1.96 & 0.00 & 0.00 & 0.0 & S & 72 \\  
&&&&&&&&&&&&&&&&&&  \\
\hline
\end{tabular}
\end{flushleft}
\end{sidewaystable*}

\addtocounter{table}{-1}
\clearpage

\begin{sidewaystable*}
\begin{flushleft}
\caption[]{The 5 GHz Catalogue (continued)}
\scriptsize
\begin{tabular}{l|ccrrrrrl|c|ccrrrrrl|c}
\hline\hline
\multicolumn{1}{c|}{} &
\multicolumn{9}{c|}{low resolution parameters} 
& \multicolumn{9}{c}{full resolution parameters}\\
\hline
 \multicolumn{1}{c|}{IAU Name}   & \multicolumn{1}{c}{R.A.}
& \multicolumn{1}{c}{DEC.} & \multicolumn{1}{c}{$S_{\rm peak}$} & 
\multicolumn{1}{c}{$S_{\rm total}$} & \multicolumn{1}{c}{$\Theta_{\rm maj}$} & 
\multicolumn{1}{c}{$\Theta_{\rm min}$} & \multicolumn{1}{c}{P.A.} &
\multicolumn{1}{c|}{} & \multicolumn{1}{c|}{$\sigma_{\rm local}$} 
&  \multicolumn{1}{c}{R.A.}
& \multicolumn{1}{c}{DEC.} & \multicolumn{1}{c}{$S_{\rm peak}$} & 
\multicolumn{1}{c}{$S_{\rm total}$} & \multicolumn{1}{c}{$\Theta_{\rm maj}$} & 
\multicolumn{1}{c}{$\Theta_{\rm min}$} & \multicolumn{1}{c}{P.A.} &
\multicolumn{1}{c|}{} & \multicolumn{1}{c}{$\sigma_{\rm local}$}       \\
 \multicolumn{1}{c|}{} & \multicolumn{2}{c}{(J2000)} & \multicolumn{2}{c}{mJy} 
& \multicolumn{2}{c}{arcsec} & \multicolumn{1}{c}{degr.} 
& \multicolumn{1}{c|}{} & \multicolumn{1}{c|}{$\mu$Jy} 
& \multicolumn{2}{c}{(J2000)}  
& \multicolumn{2}{c}{mJy} & \multicolumn{2}{c}{arcsec} 
& \multicolumn{1}{c}{degr.} & \multicolumn{1}{c|}{} 
& \multicolumn{1}{c}{$\mu$Jy}  \\
\hline
&&&&&&&&&&&&&&&&&& \\
  ATESP5 J225057-401522 & 22 50 57.79 & -40 15 22.5 & 2.89 & 3.00 & 5.61 & 0.00 & 12.8 & S & 64 & 22 50 57.79 & -40 15 22.5 & 2.41 & 2.47 & 0.00 & 0.00 & 0.0 & S & 76   \\ 
  ATESP5 J225058-401645 & 22 50 58.30 & -40 16 45.6 & 0.50 & 0.39 & 0.00 & 0.00 & 0.0 & S & 70 & 22 50 58.32 & -40 16 44.9 & 0.29 &  &  & & & D & 77  \\ 
  ATESP5 J225100-400934 & 22 51 00.94 & -40 09 34.0 & 0.49 & 0.46 & 0.00 & 0.00 & 0.0 & S & 64 & 22 51 00.96 & -40 09 34.0 & 0.31 & & & & & D & 68   \\  
  ATESP5 J225112-402230 & 22 51 12.73 & -40 22 30.1 & 1.03 & 1.13 & 0.00 & 0.00 & 0.0 & S & 71 & 22 51 12.76 & -40 22 30.2 & 0.71 & 0.85 & 1.71 & 0.85 & -14.9 & S & 70  \\  
  ATESP5 J225118-402653 & 22 51 18.34 & -40 26 53.2 & 2.64 & 2.85 & 3.26 & 2.25 & 30.0 & S & 69 & 22 51 18.40 & -40 26 53.4 & 1.40 & 2.84 & 2.96 & 2.45 & -61.5 & S & 75  \\ 
  ATESP5 J225122-402524 & 22 51 22.91 & -40 25 24.3 & 0.54 & 0.58 & 0.00 & 0.00 & 0.0 & S & 70 & 22 51 22.84 & -40 25 25.1 & 0.50 & 0.46 & 0.00 & 0.00 & 0.0 & S & 70  \\ 
  ATESP5 J225138-401747 & 22 51 38.49 & -40 17 47.3 & 0.62 & 0.64 & 0.00 & 0.00 & 0.0 & S & 66 & 22 51 38.58 & -40 17 47.5 & 0.60 & 0.74 & 2.83 & 0.00 & 5.0 & S & 71  \\  
  ATESP5 J225154-401051 & 22 51 54.97 & -40 10 51.1 & 1.15 & 1.31 & 4.95 & 0.00 & -55.3 & S & 55 & 22 51 54.96 & -40 10 51.4 & 1.18 & 1.22 & 0.00 & 0.00 & 0.0 & S & 66  \\  
  ATESP5 J225207-400720 & 22 52 07.48 & -40 07 20.6 & 4.06 & 4.12 & 0.00 & 0.00 & 0.0 & S & 57 & 22 52 07.49 & -40 07 20.6 & 3.53 & 3.66 & 0.81 & 0.10 & 25.6 & S & 65   \\  
  ATESP5 J225217-402135 & 22 52 17.19 & -40 21 35.7 & 1.73 & 1.70 & 0.00 & 0.00 & 0.0 & S & 66 & 22 52 17.20 & -40 21 35.7 & 1.54 & 1.63 & 0.95 & 0.14 & 42.5 & S & 61   \\ 
  ATESP5 J225223-401841 & 22 52 23.82 & -40 18 41.9 & 0.52 & 0.85 & 14.78 & 0.00 & -28.8 & S & 67 & 22 52 23.85 & -40 18 43.3 & 0.27 & & &  & & D & 72   \\  
  ATESP5 J225224-402549 & 22 52 24.79 & -40 25 49.1 & 4.76 & 4.74 & 0.00 & 0.00 & 0.0 & S & 55 & 22 52 24.79 & -40 25 49.3 & 4.47 & 4.60 & 0.90 & 0.00 & 3.1 & S & 67  \\   
  ATESP5 J225239-401949 & 22 52 39.29 & -40 19 49.5 & 0.68 & 1.18 & 10.06 & 7.33 & -5.9 & S & 68 & 22 52 39.35 & -40 19 47.3 & 0.31 &&&&& D($^c$) & 65   \\
  ATESP5 J225242-395949 & 22 52 42.53 & -39 59 49.9 & 0.53 & 0.42 & 0.00 & 0.00 & 0.0 & S & 63 & 22 52 42.58 & -39 59 49.6 & 0.58 & 0.51 & 0.00 & 0.00 & 0.0 & S* & 65\\   
  ATESP5 J225249-401256 & 22 52 49.92 & -40 12 56.0 & 1.09 & 0.93 & 0.00 & 0.00 & 0.0 & S & 61 & 22 52 49.90 & -40 12 56.3 & 1.01 & 1.10 & 1.81 & 0.00 & -4.7 & S & 67  \\   
  ATESP5 J225316-401200 & 22 53 16.29 & -40 12 00.6 & 0.50 & 0.52 & 0.00 & 0.00 & 0.0 & S & 63  & 22 53 16.27 & -40 12 01.1 & 0.48 & 0.54 & 0.00 & 0.00 & 0.0 & S* & 66  \\ 
  ATESP5 J225321-402317 & 22 53 21.28 & -40 23 17.7 & 0.64 & 1.21 & 28.27 & 11.48 & -53.0 & E & 67 & 22 53 21.16 & -40 23 18.1 & 0.57 & 0.59 & 0.00 & 0.00 & 0.0 & S*($^f$) & 69  \\
  ATESP5 J225322-401931 & 22 53 22.73 & -40 19 31.6 & 1.01 & 0.85 & 0.00 & 0.00 & 0.0 & S & 68 & 22 53 22.76 & -40 19 31.9 & 1.01 & 1.04 & 0.00 & 0.00 & 0.0 & S & 68  \\
  ATESP5 J225323-400453 & 22 53 23.89 & -40 04 53.7 & 0.67 & 0.85 & 6.52 & 2.56 & -44.0 & S & 59 & 22 53 23.91 & -40 04 53.6 & 0.63 & 0.64 & 0.00 & 0.00 & 0.0 & S & 68 \\ 
  ATESP5 J225325-400221 & 22 53 25.45 & -40 02 21.4 & 0.53 & 0.28 & 0.00 & 0.00 & 0.0 & S & 59 & 22 53 25.43 & -40 02 21.5 & 0.66 & 0.68 & 0.00 & 0.00 & 0.0 & S & 69   \\   
  ATESP5 J225326-395912 & 22 53 26.96 & -39 59 12.6 & 1.31 & 1.77 & 8.60 & 0.00 & 54.3 & S & 63 & 22 53 27.05 & -39 59 11.7 & 0.75 & 1.02 & 1.82 & 0.96 & 61.6 & S($^g$) & 71  \\
  ATESP5 J225332-402721 & 22 53 32.41 & -40 27 21.1 & 1.00 & 0.83 & 0.00 & 0.00 & 0.0 & S & 61 & 22 53 32.41 & -40 27 20.8 & 1.02 & 1.02 & 0.00 & 0.00 & 0.0 & S & 69  \\  
  ATESP5 J225334-401414 & 22 53 34.66 & -40 14 14.0 & 0.54 & 0.52 & 0.00 & 0.00 & 0.0 & S & 63 & 22 53 34.65 & -40 14 12.4 & 0.57 & 0.48 & 0.00 & 0.00 & 0.0 & S & 65  \\  
  ATESP5 J225344-401928 & 22 53 44.89 & -40 19 28.6 & 3.52 & 3.38 & 0.00 & 0.00 & 0.0 & S & 64 & 22 53 44.88 & -40 19 28.6 & 3.36 & 3.47 & 0.61 & 0.00 & 88.8 & S & 68  \\  
  ATESP5 J225345-401845 & 22 53 45.74 & -40 18 45.4 & 0.48 & 0.35 & 0.00 & 0.00 & 0.0 & S & 63 & 22 53 45.73 & -40 18 47.0 & 0.25 & &&&& D & 70  \\  
  ATESP5 J225353-400154 & 22 53 53.32 & -40 01 54.1 & 1.03 & 0.99 & 0.00 & 0.00 & 0.0 & S & 62 & 22 53 53.31 & -40 01 54.0 & 0.99 & 1.04 & 0.00 & 0.00 & 0.0 & S & 67 \\  
  ATESP5 J225400-402204 & 22 54 00.49 & -40 22 04.3 & 0.54 & 0.51 & 0.00 & 0.00 & 0.0 & S & 65 & 22 54 00.52 & -40 22 04.1 & 0.44 & 0.62 & 1.95 & 0.72 & -88.5 & S* & 68  \\   
  ATESP5 J225404-402226 & 22 54 04.33 & -40 22 26.9 & 3.80 & 3.74 & 0.00 & 0.00 & 0.0 & S & 63 & 22 54 04.33 & -40 22 26.7 & 3.38 & 3.78 & 1.25 & 0.57 & -17.5 & S & 71 \\  
&&&&&&&&&&&&&&&&&&   \\
\hline
\end{tabular}
\end{flushleft}
\end{sidewaystable*}

\addtocounter{table}{-1}
\clearpage

\begin{sidewaystable*}
\begin{flushleft}
\caption[]{The 5 GHz Catalogue (continued)}
\scriptsize
\begin{tabular}{l|ccrrrrrl|c|ccrrrrrl|c}
\hline\hline
\multicolumn{1}{c|}{} &
\multicolumn{9}{c|}{low resolution parameters} 
& \multicolumn{9}{c}{full resolution parameters}\\
\hline
 \multicolumn{1}{c|}{IAU Name}   & \multicolumn{1}{c}{R.A.}
& \multicolumn{1}{c}{DEC.} & \multicolumn{1}{c}{$S_{\rm peak}$} & 
\multicolumn{1}{c}{$S_{\rm total}$} & \multicolumn{1}{c}{$\Theta_{\rm maj}$} & 
\multicolumn{1}{c}{$\Theta_{\rm min}$} & \multicolumn{1}{c}{P.A.} &
\multicolumn{1}{c|}{} & \multicolumn{1}{c|}{$\sigma_{\rm local}$} 
&  \multicolumn{1}{c}{R.A.}
& \multicolumn{1}{c}{DEC.} & \multicolumn{1}{c}{$S_{\rm peak}$} & 
\multicolumn{1}{c}{$S_{\rm total}$} & \multicolumn{1}{c}{$\Theta_{\rm maj}$} & 
\multicolumn{1}{c}{$\Theta_{\rm min}$} & \multicolumn{1}{c}{P.A.} &
\multicolumn{1}{c|}{} & \multicolumn{1}{c}{$\sigma_{\rm local}$}    \\
 \multicolumn{1}{c|}{} & \multicolumn{2}{c}{(J2000)} & \multicolumn{2}{c}{mJy} 
& \multicolumn{2}{c}{arcsec} & \multicolumn{1}{c}{degr.} 
& \multicolumn{1}{c|}{} & \multicolumn{1}{c|}{$\mu$Jy} 
& \multicolumn{2}{c}{(J2000)}  
& \multicolumn{2}{c}{mJy} & \multicolumn{2}{c}{arcsec} 
& \multicolumn{1}{c}{degr.} & \multicolumn{1}{c|}{} 
& \multicolumn{1}{c}{$\mu$Jy}  \\
\hline
&&&&&&&&&&&&&&&&&&\\
  ATESP5 J225414-400853 & 22 54 14.57 & -40 08 53.0 & 1.12 & 1.95 & 9.05 & 7.88 & -40.8 & S & 64 & 22 54 14.54 & -40 08 48.8 & 0.36 & 0.84 & 9.76 & 3.90 & -16.2 & DE & 69 \\ 
&&&&&&&&&&&&&&&&&&   \\
  ATESP5 J225426-402442 & 22 54 26.16 & -40 24 42.6 & 2.06 & 4.24 & 34.73 & 0.00 & 0.0 & M   &&&&&&&&&&  \\  
  ATESP5 J225425-402446A & 22 54 25.48 & -40 24 46.9 & 2.06 & 2.19 & 3.80 & 0.00 & 52.6 & S & 64& 22 54 25.43 & -40 24 47.2 & 1.47 & 2.11 & 2.02 & 1.18 & 42.4 & S($^h$) & 70   \\
  ATESP5 J225426-402438B & 22 54 26.88 & -40 24 38.0 & 1.61 & 2.05 & 6.63 & 0.00 & 70.8 & S & 64 & 22 54 27.01 & -40 24 37.0 & 1.03 & 2.02 & 7.92 & 4.49 & 60.6 & E & 70  \\ 
&&&&&&&&&&&&&&&&&&     \\
  ATESP5 J225430-400334 & 22 54 30.47 & -40 03 34.0 & 0.63 & 0.65 & 0.00 & 0.00 & 0.0 & S & 60 & 22 54 30.47 & -40 03 34.9 & 0.58 & 0.60 & 0.00 & 0.00 & 0.0 & S & 70 \\  
&&&&&&&&&&&&&&&&&&   \\
  ATESP5 J225434-401343 & 22 54 34.70 & -40 13 43.2 & 3.17 & 7.80 & 22.51 & 3.07 & -33.6 & S & 58 & 22 54 34.35 & -40 13 37.4 & 1.59 & 3.56 & 3.06 & 2.27 & -60.5 & S($^{dh}$) & 71 \\ 
&&&&&&&&&& 22 54 35.05 & -40 13 48.9 & 1.17 & 4.38 & 6.10 & 2.51 & -38.6 & S & 
71    \\ 
&&&&&&&&&&&&&&&&&&     \\
  ATESP5 J225435-395931 & 22 54 35.26 & -39 59 31.9 & 0.71 & 0.63 & 0.00 & 0.00 & 0.0 & S & 63 & 22 54 35.29 & -39 59 32.6 & 0.69 & 0.71 & 0.00 & 0.00 & 0.0 & S & 67  \\  
  ATESP5 J225436-400531 & 22 54 36.29 & -40 05 31.3 & 0.60 & 0.65 & 0.00 & 0.00 & 0.0 & S & 65 & 22 54 36.24 & -40 05 30.8 & 0.66 & 0.68 & 0.00 & 0.00 & 0.0 & S & 66   \\  
  ATESP5 J225442-400353 & 22 54 42.64 & -40 03 53.0 & 0.56 & 0.49 & 0.00 & 0.00 & 0.0 & S & 60 & 22 54 42.63 & -40 03 53.2 & 0.54 & 0.52 & 0.00 & 0.00 & 0.0 & S & 65   \\  
  ATESP5 J225443-401147 & 22 54 43.07 & -40 11 47.4 & 0.41 & 0.39 & 0.00 & 0.00 & 0.0 & S & 60 & 22 54 43.02 & -40 11 48.1 & 0.34 &&&&& D & 71 \\  
  ATESP5 J225449-400918 & 22 54 49.73 & -40 09 18.6 & 0.86 & 0.92 & 0.00 & 0.00 & 0.0 & S & 64 & 22 54 49.72 & -40 09 18.4 & 0.72 & 0.79 & 0.00 & 0.00 & 0.0 & S & 68  \\ 
  ATESP5 J225450-401639 & 22 54 50.81 & -40 16 39.7 & 0.61 & 0.62 & 0.00 & 0.00 & 0.0 & S & 61 & 22 54 50.81 & -40 16 38.7 & 0.57 & 0.62 & 0.00 & 0.00 & 0.0 & S & 69   \\ 
  ATESP5 J225504-400154 & 22 55 04.67 & -40 01 54.8 & 3.10 & 4.59 & 10.58 & 1.82 & -40.0 & S & 58 & 22 55 04.77 & -40 01 56.3 & 1.29 & 4.58 & 16.12 & 5.35 & -31.0 & E($^i$) & 62  \\
  ATESP5 J225505-401301 & 22 55 05.25 & -40 13 01.5 & 0.37 & 0.41 & 0.00 & 0.00 & 0.0 & S & 61 &  &  & $<$0.21 &&&& & U & 70  \\ 
  ATESP5 J225509-402658 & 22 55 09.41 & -40 26 58.6 & 5.18 & 5.17 & 0.00 & 0.00 & 0.0 & S & 62 & 22 55 09.42 & -40 26 58.5 & 4.87 & 4.98 & 0.73 & 0.00 & -1.2 & S & 69 \\ 
  ATESP5 J225511-401513 & 22 55 11.94 & -40 15 13.5 & 0.58 & 0.59 & 0.00 & 0.00 & 0.0 & S & 61 & 22 55 11.96 & -40 15 13.3 & 0.55 & 0.60 & 0.00 & 0.00 & 0.0 & S & 70 \\ 
  ATESP5 J225515-401835 & 22 55 15.57 & -40 18 35.6 & 0.42 & 0.41 & 0.00 & 0.00 & 0.0 & S & 64 & & & $<$0.22 & &&& & U & 73 \\ 
  ATESP5 J225526-400112 & 22 55 26.90 & -40 01 12.1 & 0.41 & 0.23 & 0.00 & 0.00 & 0.0 & S & 64 & 22 55 26.91 & -40 01 12.0 & 0.30 & &&&& D & 67  \\ 
  ATESP5 J225529-401101 & 22 55 29.51 & -40 11 01.7 & 1.09 & 0.98 & 0.00 & 0.00 & 0.0 & S & 65 & 22 55 29.50 & -40 11 01.7 & 1.06 & 1.09 & 0.00 & 0.00 & 0.0 & S & 71 \\ 
&&&&&&&&&&&&&&&&&&  \\
\hline
\multicolumn{5}{l}{($^a$) Component B split in two sub-components.} & 
\multicolumn{6}{l}{($^d$) Source split in two components.} &
\multicolumn{5}{l}{($^g$) Hints of double structure.}\\

\multicolumn{5}{l}{($^b$) Hints of multiple structure.} & 
\multicolumn{6}{l}{($^e$) Extended structure resolved out.} & 
\multicolumn{5}{l}{($^h$) Hints of a possible core between components.}\\

\multicolumn{5}{l}{($^c$) Almost resolved out.} &  
\multicolumn{6}{l}{($^f$) Extended tail resolved out.} &  
\multicolumn{5}{l}{($^i$) Double structure catalogued as single extended 
source.} \\
\end{tabular}
\end{flushleft}
\end{sidewaystable*}

\end{document}